\documentclass[sigconf]{acmart}
\acmConference[ESEC/FSE 2022]{The 30th ACM Joint European Software Engineering Conference and Symposium on the Foundations of Software Engineering}{14 - 18 November, 2022}{Singapore}

\AtBeginDocument{
  \providecommand\BibTeX{{
    \normalfont B\kern-0.5em{\scshape i\kern-0.25em b}\kern-0.8em\TeX}}}

\usepackage{amsmath,amsfonts}
\usepackage{algorithmic}
\usepackage[ruled,linesnumbered,vlined]{algorithm2e}
\usepackage{graphicx}
\usepackage{textcomp}
\usepackage{color, colortbl}
\usepackage{xcolor}
\PassOptionsToPackage{prologue, table}{xcolor}
\usepackage{url}
\usepackage{xspace}
\usepackage{booktabs, multirow} 
\usepackage{listings}
\usepackage{tikz}
\usepackage[frozencache=true,cachedir=minted-cache]{minted} 

\usepackage{mdframed} 
\usepackage{caption}
\usepackage{subcaption}
\usepackage[normalem]{ulem}
\usepackage{diagbox}

\usepackage{wrapfig,booktabs}

\newcommand{\Comment}[1]{}
\newcommand{\Space}[1]{}
\newcommand{\parabf}[1]{\noindent\textbf{#1}}

\newcommand{\CodeIn}[1]{\begin{small}\texttt{#1}\end{small}}
\newcommand{\code}[1]{{\begin{small}\texttt{#1}\end{small}}}

\newcommand{\revision}[1]{{#1}}

\newcommand{\tech}{DeepREL\xspace} 
\newcommand{\pt}{PyTorch\xspace}
\newcommand{\tf}{TensorFlow\xspace}
\newcommand{\keras}{Keras\xspace}

\newcommand{\cradle}{CRADLE\xspace}
\newcommand{\memo}{MeMo\xspace}
\newcommand{\sbes}{SBES\xspace}

\newcommand{\numcradlemodels}{30\xspace}
\newcommand{\lemon}{LEMON\xspace}
\newcommand{\ucklee}{UC-KLEE\xspace}

\newcommand{\numlemoncoveredapis}{35\xspace}

\newcommand{\audee}{AUDEE\xspace}

\newcommand{\predoo}{Predoo\xspace}
\newcommand{\numpredoocoveredapis}{7\xspace}
\newcommand{\freefuzz}{FreeFuzz\xspace}
\newcommand{\numfreefuzzmodels}{202\xspace}
\newcommand{\fpdiff}{FPDiff\xspace}

\newcommand{\docter}{DocTer\xspace}

\newcommand{\apirelfinder}{API Matcher\xspace}
\newcommand{\progsyn}{Invocation Synthesizer\xspace}
\newcommand{\verifier}{API Match Verifier\xspace}
\newcommand{\fuzzer}{API Fuzzer\xspace}

\newcommand{\ar}{API relation\xspace}
\newcommand{\ars}{API relations\xspace}
\newcommand{\ra}{relational API\xspace}

\newcommand{\srcapi}{source API\xspace}
\newcommand{\tgtapi}{target API\xspace}

\newcommand{\NumDocAR}{2942\xspace} 
\newcommand{\NumDocEqAR}{2692\xspace}

\newcommand{\NumTotalAPI}{7973\xspace}
\newcommand{\NumTotalAPITf}{6381\xspace} 
\newcommand{\NumTotalAPIPt}{1592\xspace} 

\newcommand{\NumFreeFuzzUncoveredNewNumber}{6815\xspace}

\newcommand{\NumFreeFuzzAPI}{1158\xspace} 
\newcommand{\NumFreeFuzzAPITf}{688\xspace} 
\newcommand{\NumFreeFuzzAPIPt}{470\xspace}

\newcommand{\NumStudyAPI}{2973\xspace}
\newcommand{\NumStudyAPImore}{1815\xspace} 
\newcommand{\NumPercentStudyAPImore}{157\%\xspace}
\newcommand{\NumStudyAPIPt}{1071\xspace}
\newcommand{\NumStudyAPITf}{1902\xspace}

\newcommand{\NumVerifyPairPt}{4290\xspace}
\newcommand{\NumVerifyEqPairPt}{1357\xspace}
\newcommand{\NumVerifySimPairPt}{2933\xspace}
\newcommand{\NumVerifyPairTf}{8808\xspace}
\newcommand{\NumVerifyEqPairTf}{5132\xspace}
\newcommand{\NumVerifySimPairTf}{3676\xspace}
\newcommand{\NumVerifyEqPair}{6489\xspace}
\newcommand{\NumVerifySimPair}{6609\xspace}
\newcommand{\NumVerifyPair}{13098\xspace}
\newcommand{\NumVerifyPairSeedSeed}{1679\xspace}
\newcommand{\NumVerifyPairNewNeeded}{11419\xspace}

\newcommand{\NumAllBugs}{162\xspace}

\newcommand{\NumPreviouslyUnknownConfirm}{106\xspace}
\newcommand{\NumPreviouslyUnknownConfirmTf}{35\xspace}
\newcommand{\NumPreviouslyUnknownConfirmPt}{71\xspace}

\newcommand{\NumAllBugsPt}{121\xspace}
\newcommand{\NumAllBugsTf}{41\xspace}

\newcommand{\numFixedBugs}{29\xspace}
\newcommand{\numFixedPt}{17\xspace}
\newcommand{\numFixedTf}{12\xspace}

\newcommand{\numSeedAPI}{9\xspace}
\newcommand{\numNewAPI}{10\xspace}
\newcommand{\numSingleAPI}{19\xspace}
\newcommand{\numConsistent}{87\xspace}
\newcommand{\numConsistentSeedOnly}{35\xspace}
\newcommand{\numConsistentNew}{52\xspace}

\newcommand{\numSeedSeedPairPt}{902\xspace}
\newcommand{\numSeedSeedPairTf}{777\xspace}

\newcommand{\PortionHighPrio}{13.5\%\xspace}

\newcommand{\NumDocBugTf}{4\xspace}
\newcommand{\NumDocBugPt}{10\xspace}
\newcommand{\NumDocBug}{14\xspace}

\newcommand{\NumRejectBugTf}{1\xspace}
\newcommand{\NumRejectBugPt}{6\xspace}
\newcommand{\NumRejectBug}{7\xspace}
\newcommand{\NumOtherRejectBug}{6\xspace}

\newcommand{\NumFreeFuzzCanFindBugs}{9\xspace}

\newcommand{\NumHighPrio}{23\xspace}

\newcommand{\fpRate}{30.72\%\xspace}
\newcommand{\fpRatePt}{27.43\%\xspace}
\newcommand{\fpRateTf}{39.78\%\xspace}

\newcommand{\apione}{S}
\newcommand{\apitwo}{T}
\newcommand{\apiiter}{S'}
\newcommand{\domain}{\mathcal{D}}

\newcommand{\inp}{x}

\newcommand{\merged}{\checkmark{}}
\newcommand{\rejected}{$\times$}

\newcommand{\apipr}{API pair\xspace}
\newcommand{\apiprs}{API pairs\xspace}
\newcommand{\mapipr}{matched API pair\xspace}
\newcommand{\mapiprs}{matched API pairs\xspace}

\newcommand{\SigSim}{Signature Similarity\xspace}
\newcommand{\sigsim}{signature similarity\xspace}
\newcommand{\docsim}{document similarity\xspace}
\newcommand{\DocSim}{Document Similarity\xspace}
\newcommand{\apisig}{API signature\xspace}
\newcommand{\apisigs}{API signatures\xspace}

\newcommand{\VocabSize}{n}
\newcommand{\DocOfAPI}[1]{{{#1}}.description}
\newcommand{\sentbert}{Sentence-BERT\xspace}
\newcommand{\sbert}{{SBEncoder}}
\newcommand{\wordcnt}{c}
\newcommand{\vecone}{x}
\newcommand{\vectwo}{y}
\newcommand{\cossim}[2]{{Cos}({{#1}}, {{#2}})}
\newcommand{\simdef}[2]{{Sim}_{sig}({{#1}}, {{#2}})}
\newcommand{\simtext}[2]{{Sim}_{doc}({{#1}}, {{#2}})}
\newcommand{\maxi}[2]{{Max}({{#1}}, {{#2}})}
\newcommand{\len}{{Len}}
\newcommand{\embdef}[1]{{Rep}_{sig}({{#1}})}

\newcommand{\embtext}[1]{{Rep}_{doc}({{#1}})}

\newcommand{\APISimFunc}{{Sim_{API}}}

\newcommand{\TopK}{K}
\newcommand{\IterI}{I}
\newcommand{\NumIterTerminatePt}{8\xspace}
\newcommand{\NumIterTerminateTf}{8\xspace}

\newcommand{\invpat}{invocation code\xspace}
\newcommand{\invpats}{invocation code\xspace}

\newcommand{\lev}{{Levenshtein}}
\newcommand{\ArgsimFunc}{{Sim}_{arg}}
\newcommand{\simName}{{Sim}_{name}}
\newcommand{\simType}{{Sim}_{type}}
\newcommand{\simPos}{{Sim}_{pos}}
\newcommand{\arga}{a}
\newcommand{\argb}{b}
\newcommand{\namea}{a_{name}}
\newcommand{\nameb}{b_{name}}
\newcommand{\typea}{a_{type}}
\newcommand{\typeb}{b_{type}}

\newcommand{\idxa}{a_{idx}}
\newcommand{\idxb}{b_{idx}}

\newcommand{\argseta}{L}
\newcommand{\argsetb}{R}

\newcommand{\verinp}{verifying inputs\xspace}
\newcommand{\propa}{Equivalence$_{value}$\xspace}
\newcommand{\propb}{Equivalence$_{status}$\xspace}

\newcommand{\equivalue}{value equivalence\xspace}
\newcommand{\equistatus}{status equivalence\xspace}

\newcounter{finding}

\newcommand*\circled[1]{\tikz[baseline=(char.base)]{
            \node[shape=circle,fill,inner sep=1pt] (char) {\textcolor{white}{#1}};}}
\pdfoutput=1
\begin{document}

\title[\revision{Fuzzing Deep-Learning Libraries via Automated Relational API Inference}]{\revision{Fuzzing Deep-Learning Libraries via\\ Automated Relational API Inference}}

\captionsetup[figure]{font=bf,skip=0.2em} 
\captionsetup[table]{font=bf,skip=0.2em} 
\newcommand{\distance}{0.5em}
\setlength{\textfloatsep}{\distance} 
\setlength{\floatsep}{\distance} 
\setlength{\intextsep}{\distance} 
\setlength{\dbltextfloatsep}{\distance} 
\setlength{\dblfloatsep}{\distance}

\author{Yinlin Deng}
\authornote{Both authors contributed equally to this research.}
\affiliation{
  \institution{University of Illinois at Urbana-Champaign}
  \country{}
}
\email{yinlind2@illinois.edu}

\author{Chenyuan Yang}\authornotemark[1]
\affiliation{
  \institution{University of Illinois at Urbana-Champaign}
  \country{}
}
\email{cy54@illinois.edu}

\author{Anjiang Wei}
\affiliation{
  \institution{Stanford University}
  \country{}
}
\email{anjiang@stanford.edu}

\author{Lingming Zhang}
\affiliation{
 \institution{University of Illinois at Urbana-Champaign}
 \country{}
}
\email{lingming@illinois.edu}

\renewcommand{\shortauthors}{Yinlin Deng, Chenyuan Yang, Anjiang Wei, and Lingming Zhang}

\begin{abstract}
Deep Learning (DL) has gained wide attention in recent years. Meanwhile, bugs in DL systems can lead to serious consequences, and may even threaten human lives. As a result, a growing body of research has been dedicated to DL model testing. However, there is still limited work on testing DL libraries, e.g., \pt and \tf, which serve as the foundations for building, training, and running DL models. Prior work on fuzzing DL libraries can only generate tests for APIs which have been invoked by documentation examples, developer tests, or DL models, leaving a large number of APIs untested.
In this paper, we propose \revision{\tech{}, the first approach to automatically inferring relational APIs for more effective DL library fuzzing.} Our basic hypothesis is that for a DL library under test, there may exist a number of APIs sharing similar input parameters and outputs; in this way, we can easily ``borrow'' test inputs from invoked APIs to test other relational APIs.
Furthermore, we formalize the notion of \equivalue and \equistatus for relational APIs to serve as the oracle for effective bug finding.
We have implemented \tech as a \revision{fully automated end-to-end relational API inference and fuzzing technique for DL libraries}, which 1) automatically infers potential \ars based on API syntactic/semantic information, 2) synthesizes concrete test programs for invoking relational APIs, 3) validates the inferred relational APIs via representative test inputs, and finally 4) performs fuzzing on the verified relational APIs to find potential inconsistencies. 
Our evaluation on two of the most popular DL libraries, \pt and \tf, demonstrates that \tech can cover \NumPercentStudyAPImore more APIs than state-of-the-art \freefuzz. To date, \tech has detected \NumAllBugs{} bugs in total, with \NumPreviouslyUnknownConfirm already confirmed by the developers as previously unknown bugs. Surprisingly, \tech has detected \PortionHighPrio of the high-priority bugs for the entire \pt issue-tracking system in a three-month period. Also, besides the \NumAllBugs{} code bugs, we have also detected \NumDocBug documentation bugs (all confirmed).  
\end{abstract}

\maketitle

\section{Introduction}
\label{section:introduction}
Recent years have witnessed the surge of deep learning (DL) in a variety of applications, including computer vision~\cite{He_2016_CVPR,simonyan2014very}, natural language processing~\cite{graves2013speech,gers2000learning}, robotics~\cite{eitel2015multimodal,lenz2015deep}, bioinformatics~\cite{ravi2016deep,min2017deep}, and software engineering~\cite{yang2020survey,li2019deepfl,goffi2016automatic,li2020dlfix,mesbah2019deepdelta,gu2018deep,white2016deep,ye2021automated, zeng2022study, zeng2021deep}. Meanwhile, similar as traditional software systems, DL systems can also have bugs, which can lead to serious consequences and may even threaten human lives~\cite{uberkill}. 

To date, most prior work on DL testing focused on testing/verifying DL models, with an emphasis on adversarial attacks~\cite{goodfellow2014explaining,moosavi2016deepfool,carlini2019evaluating,madry2017towards,akhtar2018threat,papernot2016limitations}, metrics for model testing~\cite{ma2018deepgauge,yan2020correlations,harel2020neuron,pei2017deepxplore,kim2019guiding}, application-specific model testing~\cite{zhang2018deeproad,zhou2020deepbillboard,tian2018deeptest}, and verifying certain properties of models~\cite{liu2019algorithms,amir2020smt}. Meanwhile, there is limited work targeting the reliability of DL libraries, which serve as the central infrastructures for building DL models, and are the foundation for running, optimizing and deploying DL models. \cradle~\cite{cradle} is the trailblazing work for testing DL libraries, which resolves the oracle challenge with differential testing of various DL models on multiple backends of \keras~\cite{keras}. \audee~\cite{guo2020audee} and \lemon~\cite{lemon} further augment \cradle by leveraging search-based mutation strategies to generate more diverse DL models/inputs for testing library code.
Different from the above model-level DL library testing techniques, more recently, \freefuzz~\cite{freefuzz} has been proposed to mine example inputs from open source (including code snippets from the library documentation, developer tests, and DL models in the wild) to directly test each DL library API in isolation. \freefuzz has been evaluated on \pt~\cite{paszke2019pytorch} and \tf~\cite{abadi2016tensorflow}, currently the two most popular DL libraries (with 54K/162K stars on Github). The experimental results show that \freefuzz can cover $9\times$ more APIs than state-of-the-art \lemon and detect various previously unknown bugs.

Despite the promising results, existing techniques on fuzzing DL libraries still suffer from the following limitations. First, the input generation is still far from optimal. \cradle and \audee can only test APIs that are covered in the original models, and \lemon can cover slightly more APIs with layer mutations; furthermore, although \freefuzz can cover up to \NumFreeFuzzAPI APIs for \pt and \tf (which is already a huge improvement over other work), it is still unable to test an API if there is no code snippet directly invoking the API.
Second, there is still a lack of powerful test oracles. Existing techniques typically perform differential testing across different DL libraries or hardware backends (e.g., GPU/CPU) to address the test oracle issue. However, differential testing across DL libraries is typically applied at the model level and suffers from the limited effectiveness of model-level testing (e.g., limited API coverage and accumulated floating-point precision loss)~\cite{cradle, lemon, guo2020audee}, while different backends often share common code logic/design (and thus may also share similar bugs)~\cite{freefuzz}. Thus, it is also crucial to investigate novel test oracles for effective DL library fuzzing.

To address the aforementioned limitations, in this work, we open a new dimension for \revision{testing DL libraries via automated relational API inference. The inspiration stems from the fact that prior work~\cite{memo, goffi2014search,mattavelli2015synthesis, vanover2020discovering} has discovered a number of equivalent APIs in traditional software systems (e.g., Java projects)\footnote{\revision{Some of such existing work~\cite{goffi2014search, mattavelli2015synthesis, memo}) treated the entire software systems under test as the test objects, and thus viewed this as \emph{metamorphic testing}~\cite{chen2020metamorphic}. In this paper, we treat each API as a test object and view this as \emph{differential testing} (following ~\cite{vanover2020discovering}).}}.
We envision such relational API inference also to be an inspiring direction for fuzzing DL libraries. In this way, given the same inputs generated via fuzzing, APIs that are equivalent in functionality should produce the same numerical results (i.e., \emph{\equivalue}). Moreover, besides the previously studied equivalent APIs, we further leverage the fact that DL APIs with similar functionality should behave similarly in terms of program status (i.e., \emph{\equistatus}) for more effective fuzzing. For example, although \code{torch.nn.AdaptiveAvgPool3d} and \code{torch.nn.AdaptiveMaxPool3d} in \pt are not equivalent, they are functionally similar APIs; thus, we can feed any valid input of the first API to the second API and expect its invocation to also be successful.}
Based on this intuition, we can easily ``borrow'' test inputs generated for one API to test other \ra{}s. Also, API relations can directly serve as test oracle for differential testing. Therefore, we can easily overcome the aforementioned limitations.

We have built a fully-automated technique, \revision{ \tech, which infers such API relations without human intervention for fuzzing DL libraries.} One key challenge is how to obtain the API relations automatically and accurately. Existing work~\cite{goffi2014search, mattavelli2015synthesis, memo, vanover2020discovering} on equivalent API inference for traditional software systems can hardly be applied for DL library testing, e.g., the most recent MeMo work~\cite{memo} heavily relies on well-documented API relations, which are rare in DL libraries. To this end, \tech first automatically infers all possible candidate \mapiprs based on API syntactic and semantic information. Then, \tech synthesizes concrete test programs for those potentially relational APIs. After that, \tech leverages a set of representative valid inputs (automatically traced during prior normal API executions) to check whether the inferred API relations hold or not. Lastly, \tech takes the validated API pairs, and leverages mutation-based fuzzing to generate a much diverse and extensive set of test inputs for detecting potential inconsistencies among relational APIs. Our study has shown for the first time that there can be a surprising number of equivalent or similar APIs within popular DL libraries (e.g., \NumVerifyPairPt/\NumVerifyPairTf verified relational API pairs by \tech for \pt/\tf), which can substantially help with fuzzing
DL libraries (and beyond).%
In summary, our paper makes the following contributions:
\begin{itemize}
    \item \textbf{Dimension.} This paper opens a new dimension for \revision{fully-automated DL library fuzzing via relational API inference.}
    \item \textbf{Technique.} We build \tech, a \emph{fully-automated} end-to-end framework for DL library testing. \tech automatically infers all possible candidate relational APIs based on both API syntactic and semantic information, and then dynamically verifies them via test program synthesis. While this work focuses on DL libraries, the basic idea of \tech is general and can also be applied to other software systems.%
    \item \textbf{Evaluation and Impact.} \tech covers \NumStudyAPImore more APIs than prior work (i.e., \NumPercentStudyAPImore improvement), and has detected \NumAllBugs{} bugs in total, with \NumPreviouslyUnknownConfirm already confirmed by the developers as previously unknown bugs. Surprisingly, \tech was able to detect \PortionHighPrio of the high-priority bugs for the entire \pt issue-tracking system in a three-month period. Also, besides the \NumAllBugs{} code bugs, we were also able to detect \NumDocBug documentation bugs (all confirmed) as a by-product of our experimentation.
\end{itemize}

\section{Background}
\label{section:background}
\begin{figure}[t]
    \centering
    \includegraphics[keepaspectratio=true,width=\columnwidth]{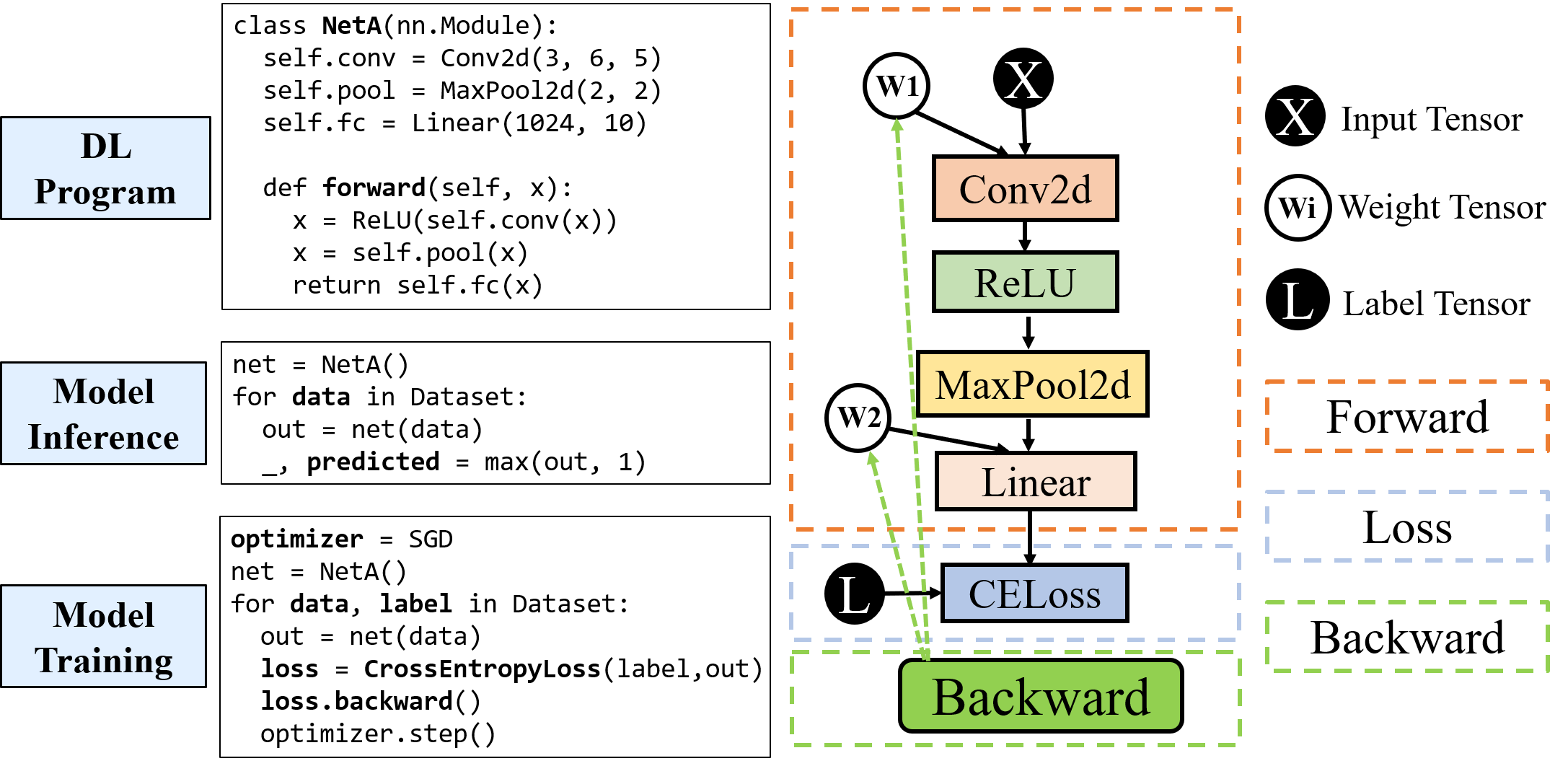}

    \caption{Background knowledge on DL Models and APIs}
    \label{fig:background}
\end{figure}

\subsection{Basics about DL Models and APIs}
\label{subsec:basics}
For DL models, \emph{inference} is the process of using a fixed DL model to complete a specific task for unseen data, while \emph{training} is the process for a neural network to update its weights to learn how to better perform a certain task given the labeled data (under the scenario of supervised learning~\cite{wikisupervisedlearning}). We next shed light on how this is achieved by APIs from DL libraries and the principles behind.

\parabf{DL Models.}
To build and run a DL model, developers first need to define the model by writing a DL program in deep learning libraries (e.g., \pt~\cite{paszke2019pytorch}  and \tf~\cite{abadi2016tensorflow}). Take the DL program \CodeIn{NetA} written in \pt (shown on the left-hand side of Figure~\ref{fig:background}) as an example, it includes a convolution layer (\CodeIn{Conv2d}), a max pooling layer (\CodeIn{MaxPool2d}), and a linear layer (\CodeIn{Linear}). The function \CodeIn{forward} defines how the input tensor (\CodeIn{x}) should flow in the defined layers (and other related APIs). Besides input tensors, there are also \emph{weight tensors} (e.g., \CodeIn{W1} and \CodeIn{W2} shown on the right-hand side of Figure~\ref{fig:background}). Their values will be updated during training, the process of which is called \emph{back-propagation}~\cite{wikibp}, a procedure natively supported by DL libraries. Training the model is achieved by first running the forward part (i.e., inference) of the neural network (\CodeIn{out = net(data)}), computing the loss (\CodeIn{loss = CrossEntropyLoss(label, out)}), computing the gradient (\CodeIn{loss.backward()}), and invoking the optimizer (\CodeIn{optimizer.step()}) for back-propagation.

\parabf{DL APIs.}
When running a DL model, the APIs involved in building the model are also executed. 
\revision{Essentially, writing a DL program  can be viewed as defining a \emph{computation graph}.} It is a directed acyclic graph (DAG) whose nodes stand for DL APIs while edges represent the flow of the tensors.
Figure~\ref{fig:background} also shows the computation graph for the example neural network. It composes of three parts: forward part (taking input tensors and weight tensors as input), loss computation part (requiring label tensors), and backward part (for updating weight tensors). Actually the backward part needs to construct a much more complex graph~\cite{zheng2021neoflow}, but we omit it in Figure~\ref{fig:background} for simplicity. Essentially, running a whole DL model can be broken down into invoking a series of DL APIs based on their topological sorting~\cite{toposort} of the computation graph.

\subsection{Fuzzing DL Libraries}
\label{subsec:fuzzingDL}
To our knowledge, there are mainly two categories of work for fuzzing DL libraries, model-level testing and API-level testing.

\parabf{Model-level Testing.}
\cradle~\cite{cradle} is the first to apply differential testing for DL libraries. Given the fact that \keras~\cite{keras} is a library featuring high-level APIs for building DL models, APIs in \keras may have multiple implementation in its supported lower-level libraries. Therefore, \cradle takes \numcradlemodels pre-trained DL models as input, and runs differential testing to find inconsistencies between different low-level libraries for \keras.
 \revision{More recently, \audee~\cite{guo2020audee} and \lemon~\cite{lemon} have proposed to use search-based mutation strategies to generate mutated DL models for differential testing on different backends}. While \audee focuses on mutating parameters of layers, weight tensors, and input tensors, \lemon applies mutation rules by adding/deleting layers and changing the values of weight tensors.
In this way, \lemon's mutation rules are more general, and can cover more APIs than the original DL models.
However, even \lemon's model-level mutation can only be applied to a limited number of APIs given a DL model.
For instance, the intact-layer mutation rule~\cite{lemon} proposed in \lemon requires that the output tensor shape of the API to be added to (or deleted from) the model should be identical to its input tensor shape. This constraint makes a large number of APIs inapplicable for model-level mutation. Researchers have recently shown that \lemon can hardly invoke additional library code or APIs with its mutation rules~\cite{freefuzz}.
In general, model-level testing suffers from covering only a limited number of APIs, e.g., even state-of-the-art \lemon can only cover \numlemoncoveredapis APIs for \tf.

\parabf{API-level Testing.}
Different from prior work on DL library fuzzing, the recent \freefuzz work~\cite{freefuzz} proposes to directly mine test inputs from open source for  API-level fuzzing, a much finer grained level than model-level testing. \revision{One challenge is that Python is a dynamically-typed language, and thus it is hard to determine the types of API parameters for fuzzing Python APIs.} Prior work has to manually set up the API arguments, and thus can only test a small number of APIs/operators, e.g., \predoo~\cite{zhang2021predoo} can only test \numpredoocoveredapis APIs for \tf . \revision{A very recent work \docter~\cite{docter} constructs rules to extract DL-specific input constraints from API documentation and uses the constraints to generate valid/invalid inputs for testing DL libraries; meanwhile, it requires manual annotation for 30\% of API parameters.}
In contrast, \freefuzz resolves this challenge fully automatically via dynamically tracing API executions in code snippets from documentation, developer tests, and \numfreefuzzmodels DL models. More specifically, \freefuzz records the traced argument values in a database, and further performs mutation-based fuzzing to mutate those traced values to generate even more inputs for fuzzing DL library APIs.
 Lastly, \freefuzz applies differential testing on different hardware backends (i.e., CPU/GPU) for detecting potential consistency bugs.
Despite its big improvement over prior work, \freefuzz can only test \NumFreeFuzzAPI APIs for \pt and \tf, which are the ones covered in its input mining stage, \revision{leaving a total of \NumFreeFuzzUncoveredNewNumber APIs uncovered.} Also, different hardware backends may still share code logics/design, causing the differential testing oracle used by \freefuzz to miss various bugs. In this work, we propose to test \ra{}s to further overcome such limitations. \revision{We build our technique (\tech) upon \freefuzz to automatically infer relational APIs and leverage them to fuzz DL libraries. Note, however, that our \tech idea is general and can be built on any other API-level fuzzer for DL libraries (e.g., \docter~\cite{docter}). We choose \freefuzz since it is a recent state-of-the-art technique that is both publicly available and fully automated.}

\newcommand{\dl}{DL\xspace}
\newcommand{\api}{\mathcal{A}}
\newcommand{\stat}[1]{\llbracket #1 \rrbracket}

\newcommand{\status}{status\xspace}
\newcommand{\statuses}{statuses\xspace}
\newcommand{\crash}{\texttt{Crash}\xspace}
\newcommand{\exception}{\texttt{Exception}\xspace}
\newcommand{\success}{\texttt{Success}\xspace}

\section{Preliminaries}
\label{subsec:preliminary}

We first introduce the preliminaries for our fuzzing technique in this section. Given the set of all possible APIs, $\api$, for a \dl library under test, we aim to define the relational property between the invocation results of a source API $\apione\in \api$ and a target API $\apitwo\in \api$.

\begin{figure}
    \begin{minted}[mathescape, linenos, numbersep=5pt, gobble=0, fontsize=\footnotesize, frame=lines, framesep=2mm]{python}
result1 = torch.broadcast_shapes(*shapes)
result2 = torch.broadcast_tensors(*map(torch.empty, shapes))[0].shape
    \end{minted}
	\caption{API pair with the same output}
	\label{fig:valuematch}
\end{figure}

Intuitively, we can directly check whether $\apione$ and $\apitwo$ produce equivalent outputs. For example, Figure ~\ref{fig:valuematch} shows an API pair which, according to the \pt documentation ~\cite{ptwebdocbroadcast}, should always produce the same results. The \code{torch.broadcast\_shapes} API applies broadcasting on a list of compatible shapes to align them. The \code{torch.broadcast\_tensors} API applies broadcasting on a list of shape-compatible tensors to align their shapes. In fact, the first API can be rewritten as 1) creating intermediate empty tensors from tensor shapes with \code{map} and \code{torch.empty}, 2) calling \code{torch.broadcast\_tensors} with these tensors, and 3) getting the shape of the output tensor. Since the source and target APIs can achieve the same functionality with totally different implementations \revision{given the same input (\CodeIn{shapes} in Figure~\ref{fig:valuematch})}, they provide a great opportunity for differential testing. Therefore, we have the following formal definition:
\begin{definition} 
\textbf{\propa.} Given a set of inputs $\domain$, source API $\apione\in \api$ and target API $\apitwo\in \api$ satisfy \propa{} (modulo $\domain$) iff their invocations always output the same results given any input in $\domain$. Formally,
\begin{equation}
    S  \equiv T (mod \  \domain) \iff \forall \inp \in \domain.~ S(x) = T(x) \quad 
\end{equation}
\end{definition}

\begin{figure}
    \begin{minted}[mathescape, linenos, numbersep=5pt, gobble=0, fontsize=\footnotesize, frame=lines, framesep=2mm]{python}
layer1 = torch.nn.AdaptiveAvgPool3d(output_size)
result1 = layer1(input)
layer2 = torch.nn.AdaptiveMaxPool3d(output_size)
result2 = layer2(input)
    \end{minted}
	\caption{API pair with different outputs but same status}
	\label{fig:statusmatch}
\end{figure}

While this can be effective in detecting potential consistency bugs, the checking is too strict and may not apply to a large number of APIs. In fact, it could be possible that $\apione$ and $\apitwo$ produce totally different results, but tend to behave similarly given similar inputs. For example, the \apipr shown in Figure~\ref{fig:statusmatch} does not hold the \propa property since the output of \code{AdaptiveAvgPool3d} is different than \code{AdaptiveMaxPool3d}. The first API applies a 3D adaptive \textit{average} pooling over an input but the latter applies a 3D adaptive \textit{maximum} pooling. 
However, these two APIs do have something in common in terms of functionality that both of them apply a pooling operation, which is also valuable for testing. 
 Therefore, we further abstract the invocation results of a program into a set of coarse-grained \emph{\statuses}: \success, \exception, and \crash. \success denotes that program executions terminate normally, while \exception means that program executions throw known exceptions. Lastly, \crash represents the cases where the program executions crash with unexpected errors, e.g., segmentation faults or \textit{\code{INTERNAL ASSERT FAILED} } errors (which are ``\textit{never acceptable}'' as commented by \pt developers). We then further introduce the notation of $\stat{\cdot}\in \{\text{\success, \exception, \crash}\}$ to return the execution status of the input program. For example, $\stat{\apione(\inp)}=\text{\success}$ indicates that $\apione$ terminates normally with input $\inp$. In this way, we can define another property for checking potential consistency:

\begin{definition} 
\textbf{\propb.} Given a set of inputs $\domain$, source API $\apione\in \api$ and target API $\apitwo\in \api$ satisfy \propb{} (modulo $\domain$) iff their invocation always output the same statuses given any input in $\domain$. Formally,
\begin{equation}
    S  \sim T (mod \  \domain) \iff \forall \inp \in \domain.~ \stat{S(x)} = \stat{T(x)} \quad 
\end{equation}
\end{definition}

To conclude, the \propb{} relation is a relaxed notation for the \propa{} relation, which is in turn a relaxed notation of semantic equivalence (denoted as $S  \equiv T$). Formally, 
\begin{equation}
    S  \equiv T~\Longrightarrow~   S  \equiv T (mod \  \domain)~\Longrightarrow~ S  \sim T (mod \  \domain)  \quad 
\end{equation}

One crucial component of these two definitions is the domain $\domain$ on which the properties are constrained on. Aiming for more accurate API relations, it would be beneficial to cover more representative test inputs within the intersection of the valid input space of the source and target APIs as the domain.

\section{Fuzzing Relational APIs}
\label{section:approach}

\begin{figure*}[htb]
    \captionsetup{justification=centering}
    \centering
    \includegraphics[keepaspectratio=true,width=\textwidth]{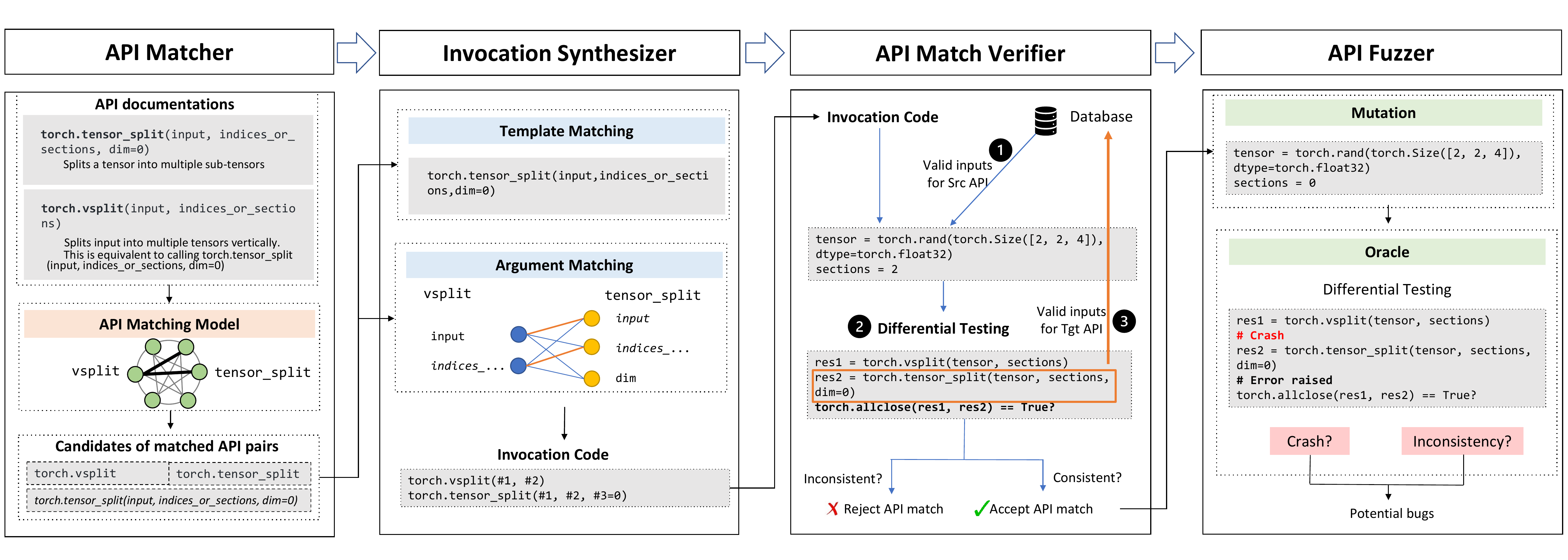}
    \caption{\tech overview} 
    \label{fig:overview}
\end{figure*}

Figure~\ref{fig:overview} shows the overview of our \tech{} technique for fuzzing relational APIs of DL libraries. \tech{} takes as input the targeted DL library, its API documentation, and a database of valid historical API invocations (e.g., automatically collected via running documentation examples, library tests, and DL models~\cite{freefuzz}). Each entry of the database contains the concrete argument values passed into an API during invocation, and is obtained through dynamic tracing. Overall, \tech{} performs the following four phases iteratively:

\emph{API Matcher (Section~\ref{subsec:apirelafinder}).} In order to test a DL library which typically has hundreds or even thousands of APIs, the first challenge is to identify the API pairs that are likely to satisfy the desired properties \propa or \propb. \apirelfinder maps each API into embeddings based on API documentation, and uses embedding similarity to identify  candidate matched APIs. 

\emph{Invocation Synthesizer (Section~\ref{subsec:progsyn}).} Given a collection of potential matched API pairs, \progsyn{} decides how to invoke them. To construct valid invocations for later verification, we impose a constraint on the source API: it must have at least one (valid) invocation in the database. In this way, given an invocation of the \srcapi, \progsyn aims to synthesize the invocation code for the \tgtapi. 

\emph{API Match Verifier (Section ~\ref{subsec:verifier}).} Given the invocation code of matched APIs, this phase would check whether each API pair satisfies property \propa or \propb with a set of representative inputs as the verifying test inputs.  If the result values (resp. execution statuses) are consistent for all tests, then \verifier accepts the \apipr as \propa (resp. \propb). If \verifier detects any inconsistency in this phase, it then rejects the API pair. 

\emph{API Fuzzer (Section ~\ref{subsec:fuzzing}).} The last step is to leverage the verified API pairs to detect potential consistency bugs. \fuzzer uses mutation-based fuzzing to generate a large number of test inputs for \srcapi{}s, and tests the verified \apiprs with oracles (Section ~\ref{subsec:preliminary}). 

Lastly, recall that in order to generate valid inputs, the \srcapi{} must have at least one (valid) invocation in the database. \tech further adopts an iterative process to cover more API pairs (Section~\ref{subsec:loop}). The newly generated valid \tgtapi invocations can be added to the database to serve as the \srcapi{}s for the next iterations to detect more potential API pairs. 
The following sections would explain each phase in detail.

\subsection{\apirelfinder}
\label{subsec:apirelafinder}

In this phase, \tech identifies potential \mapiprs from documentation.   \tech{} uses \apirelfinder to infer similar API pairs as \mapipr candidates (which will be further verified later). \apirelfinder would map each API into API embeddings, and compute similarities of each \apipr to be the distance of their embeddings. We consider \emph{\sigsim} and \emph{\docsim} that cover both API syntactic and semantic information for similarity computation.
Overall, for each API pair, the similarity is defined as the maximum of the two:
\begin{equation}
    \APISimFunc(\apione, \apitwo) = \maxi{ \simdef{\apione}{\apitwo}}{\simtext{\apione}{\apitwo} }
\end{equation}
We compute the pair-wise similarity for every API pair, and pair each API with its $\TopK$-closest neighbours as the candidate \mapiprs. $\TopK$ is a hyper-parameter and it is set to $10$ in the default setting of \tech{}.  Notably, we also analyze the impact of different values of $\TopK$ in our experimental study (Section~\ref{sec:rq3}).

\parabf{\SigSim.} 
 The signature of an API contains the API name and an ordered list of argument names. The APIs to be paired tend to follow a similar syntactic pattern in terms of their signature. For example, \code{tf.math.maximum} and \code{tf.math.minimum} are two APIs satisfying \propb, and their signatures are very similar: \code{tf.math.maximum(x, y, name=None)} and \code{tf.math.minimum(x, y, name=None)}. We map an API signature into its TF-IDF (term frequence - inverse document frequency) embedding ~\cite{enwiki:1071253989}, and use the embedding distance as the similarity measure.

TF-IDF has been widely adopted in the field of information retrieval, and it reflects the importance of each word in a document. Some common words like \code{tf} and \code{torch} in the \apisig are less informative, so their TF-IDF weights tend to be smaller. To obtain the TF-IDF embedding for each API, we first break the \apisigs into subwords (also called tokens) and then standardize them. Let $\VocabSize$ denote the size of the vocabulary from all the tokenized \apisigs, an API $\apione$ can be represented as an unnormalized term frequency embedding $[\wordcnt^\apione_1, \wordcnt^\apione_2, \ldots, \wordcnt^\apione_{\VocabSize}]$, where $\wordcnt^\apione_j$ is the number of occurrences of word $j$ in the \apisig of $\apione$. We further normalize it with the inverse document frequency for each word to get the TF-IDF embedding:
\begin{equation}
    \embdef{\apione} = [\frac{\wordcnt^\apione_1}{\sum\limits_{\apiiter \in \api}\wordcnt^{\apiiter}_1}, \frac{\wordcnt^\apione_2}{\sum\limits_{\apiiter \in \api}\wordcnt^{\apiiter}_2}, \ldots, \frac{\wordcnt^\apione_\VocabSize}{\sum\limits_{\apiiter \in \api}\wordcnt^{\apiiter}_\VocabSize}]
\end{equation}

The cosine similarity of two vectors is the cosine of the angle between them, and thus always belongs to the interval $[-1, 1]$. The cosine similarity of two arbitrary vectors $\vecone, \vectwo$ is defined as follows:
\begin{equation}
    \cossim{\vecone}{\vectwo} = \frac{\vecone \cdot \vectwo}{ \| \vecone\| \| \vectwo \|}
\end{equation}

We then compute the cosine similarity between the TF-IDF embeddings to be the \sigsim of two APIs $(\apione, \apitwo)$:
\begin{equation}
    \simdef{\apione}{\apitwo} = \cossim{\embdef{\apione}}{ \embdef{\apitwo}}
\end{equation}

\parabf{\DocSim.}
To complement the signature similarity, we further model the semantic similarity between API documents. We extract API descriptions from API documents, each of which is a one-sentence summary of an API given in the beginning of the document. For example, API \code{torch.vsplit} is described as ``\emph{Splits input, a tensor with two or more dimensions, into multiple tensors vertically according to indices\_or\_sections. }'' The description succinctly and surgically states the expected input, the transformation applied, and the expected output.
We use \sentbert ~\cite{reimers2019sentence} to encode these informative description sentences into semantically meaningful sentence embeddings. \sentbert targets specifically on generating embeddings whose cosine distance reflects the semantic textual similarity. We first collect all API descriptions from the documentation. The description for an API $\apione$ is denoted as $\DocOfAPI{\apione}$. We use $\sbert$ to denote the \sentbert encoder, which takes a natural language sentence as input, and outputs a vector in high-dimensional space. For each API $\apione$, we encode it with $\sbert$ to obtain its document embedding:

\begin{equation}
    \embtext{\apione} = \sbert(\DocOfAPI{\apione})
\end{equation}

For each API pair $(\apione, \apitwo)$, we compute the cosine similarity between their document embeddings as their \docsim:
\begin{equation}
    \simtext{\apione}{\apitwo} = \cossim{ \embtext{\apione}}{\embtext{\apitwo}} \quad 
\end{equation}

\subsection{\progsyn}
\label{subsec:progsyn}

In this phase, we leverage argument matching and template matching to synthesize the \invpat for each \mapipr. Note that the invocation code of \srcapi is simply the code snippet that directly invokes the \srcapi with the valid traced inputs. Therefore, we will next focus on generating the target API invocation code.

\parabf{Argument Matching.} 
For each \mapipr candidate, \tech first synthesizes the \invpat based on API definitions. It maps the arguments of the source API to the arguments of the target API to synthesize the invocation code of the \tgtapi (with the arguments from the \srcapi).

We transform the argument matching problem into a \emph{maximum weighted bipartite matching} problem~\cite{enwiki:1033788761}. \tech generates the \invpat based on the best argument match. 
More formally, given the \srcapi $\apione$, the \tgtapi $\apitwo$, and their argument lists $\apione.args$ and $\apitwo.args$, the corresponding bipartite graph is $G=(\argseta, \argsetb, E)$, where $\argseta=\{\arga \mid \arga \in \apione.args\}$, $\argsetb=\{\argb \mid \argb \in \apitwo.args\}$ and $E=\{(\arga, \argb) \mid \arga \in \argseta, \argb \in \argsetb\}$. The weight of each edge $(\arga, \argb) \in E$ is the similarity $\ArgsimFunc(\arga, \argb)$ of $\arga$ and $\argb$, which is defined as:
\begin{equation}
    \ArgsimFunc(\arga, \argb) = \simName(\arga, \argb) + \simType(\arga, \argb) + \simPos(\arga, \argb)
\end{equation}
The similarity $\ArgsimFunc(\arga, \argb)$ is determined by the names, potential types, and positions of the arguments. First, the similarity of two argument names is computed based on the following formula:
\begin{equation}
    \simName(\arga, \argb) = 1 - \frac{\lev(\namea, \nameb)}{\maxi{\len(\namea)}{\len(\nameb)}}
\end{equation}
where $\namea$ is the name of argument $\arga$. 
This is based on the Levenshtein Distance~\cite{enwiki:1067984406} between two names. Next we compute the similarity of two type sets as:
\begin{equation}
    \simType(\arga, \argb) = \frac{|\typea \cap \typeb|}{|\typea|}
\end{equation}
where $\typea$ is the set of possible types of $\arga$. 
If the set of types that argument $\argb$ can take contains all possible types of $\arga$, $\arga$ is more likely to be mapped to $\argb$ since all types of $\arga$ are legal for $\argb$. 

We also compute the positional similarity of two arguments as:
\begin{equation}
    \simPos(\arga, \argb) = 1 - \frac{|\idxa - \idxb|}{\maxi{\len(\apione.args)}{\len(\apitwo.args)}}
\end{equation}
where $\idxa$ is the index of $\arga$ in $\apione$'s argument list. For example, if $\arga$ and $\argb$ are both the first argument for $\apione$ and $\apitwo$, then $|\idxa - \idxb|$ equals to $0$ and thus their positional similarity is $1$.

After constructing this graph, \tech leverages the Kuhn–Munkres algorithm~\cite{munkres1957algorithms} to find the best argument match and synthesizes the invocation code based on it.  When the source and target APIs have the same number of arguments, \tech will synthesize the \invpat based on the best argument match directly. Otherwise, when they have different numbers of arguments, if the unmatched arguments contain non-optional arguments, argument matching will abort for the current API pair since the search space for determining the values of those unmatched non-optional arguments is huge. That said, \tech only considers the case where the optional arguments of the source or target API are not matched. For the unmatched optional arguments, \tech just uses their default values (Python optional arguments always have default values).

\begin{figure}
    \centering
    \includegraphics[keepaspectratio=true,width=0.9\columnwidth]{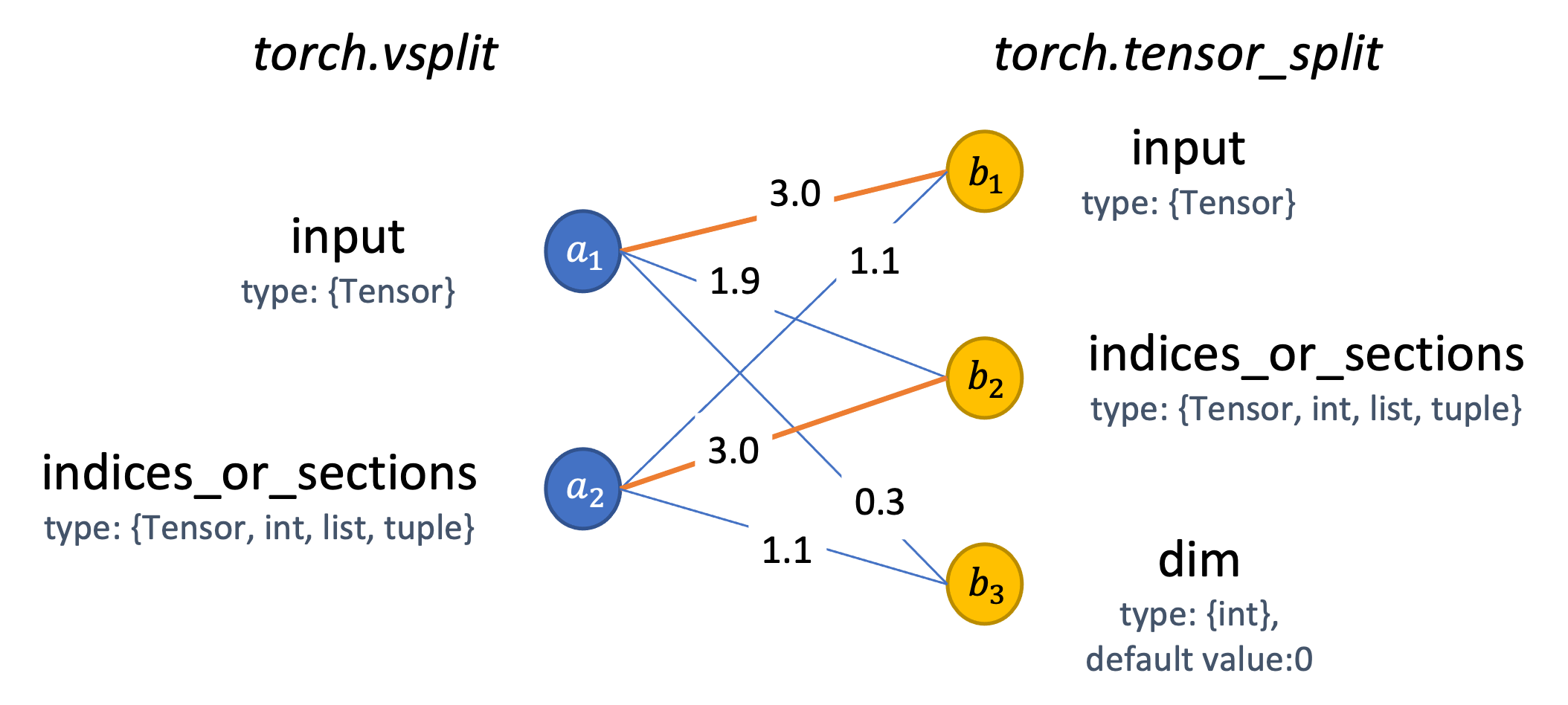}
    \caption{Example weighted bipartite graph}
    \label{fig:argmatch}
\end{figure}

For instance, consider API pair \code{torch.vsplit} and \code{torch.tensor\_split}. The corresponding weighted bipartite graph is shown in Figure~\ref{fig:argmatch}. 
For each vertex, its name and type information are marked next to it, such as vertex $a_1$, whose argument name is \code{input}, possible type set is composed of \code{Tensor}, and index is 1. 
The weight of each edge, that is, the similarity of the two arguments, is marked on the corresponding edge.
For vertices $a_1$ and $b_1$, because they have exactly the same name, possible type and index, the similarity between them is $3$. For vertices $a_1$ and $b_2$, their names are different but the type that $a_1$ can take is legal for $b_2$, so they have a relatively high similarity of $1.9$. However, for $a_2$ and $b_1$, only one type of $a_2$ is legal for $b_1$, which causes them to have a low similarity of $1.1$. Overall, the best match in the graph is $\{(a_1, b_1), (a_2, b_2)\}$, leaving $b_3$, the optional argument \code{dim} of \code{torch.tensor\_split}, unmatched. Then \tech will set \code{dim} as its default value $0$ to generate the \invpat, as shown in Figure~\ref{fig:enumeratematch} (where placeholders \#$i$ indicate the argument mapping between source and target APIs).

\begin{figure}[t]
	\begin{lstlisting}[basicstyle=\ttfamily\footnotesize,escapeinside={(*@}{@*)},columns=fixed,xleftmargin=3.5ex,language=Python]
torch.vsplit(#1, #2)
torch.tensor_split(#1, #2, dim=0)
    \end{lstlisting}
	\caption{Invocation synthesis via argument matching  }
	\label{fig:enumeratematch}
\end{figure}

\begin{figure}
\captionsetup{justification   = raggedright,
              singlelinecheck = false}
\centering
\begin{subfigure}{\columnwidth}
    \includegraphics[width=\textwidth]{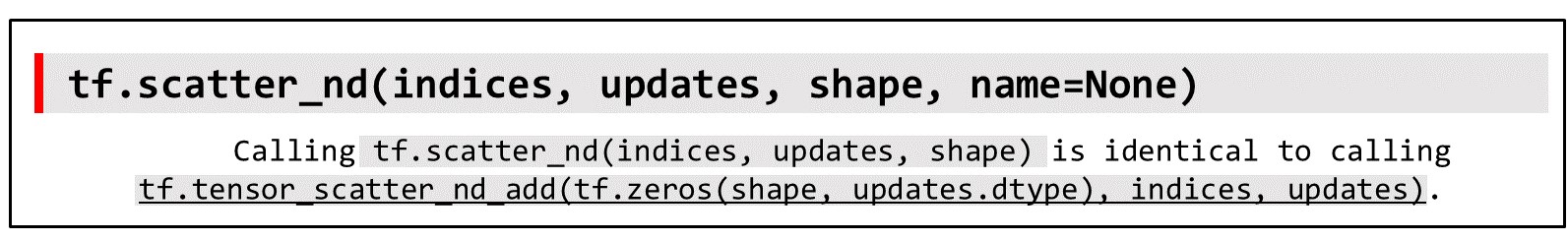}
    \caption{Documentation of \code{tf.scatter\_nd}}
    \label{fig:templatedoc}
\end{subfigure}
\hfill

\begin{subfigure}{\columnwidth}
    \begin{lstlisting}[basicstyle=\ttfamily\footnotesize,escapeinside={[@}{@]},columns=fixed,xleftmargin=3.5ex, language=Python]
tf.scatter_nd(#1, #2, #3)
tf.tensor_scatter_nd_add(tf.zeros(#3, #2.dtype), #1, #2)
    \end{lstlisting}
	\caption{Invocation code from template}
	\label{fig:matpat}
\end{subfigure} 
\caption{Invocation synthesis via template matching }
\label{fig:broadcasttemplate}
\end{figure}

Note that for every argument of every API, we gather the possible types that it can take in the invocation traces from open source. 
To be precise, we extract the argument type information from all the invocation cases of the API to form the possible type set. The type set depends on the traces and therefore is likely to be incomplete. When no traces cover a particular argument $\argb$ of a target API, the type set $\typeb$ is an empty set; thus, the type similarity between $\argb$ and any other argument $\arga$ will be $\simType(\arga, \argb)=0$. Note that the argument matching algorithm still works in this scenario, with only the name and positional similarities being considered.
   \Comment { }

\parabf{Template Matching.} 
For some complex \mapiprs, \tech cannot leverage argument matching to generate the correct invocation code of \tgtapi. 
For example, Figure~\ref{fig:matpat} shows the right \invpat between \code{tf.scatter\_nd} and \code{tf.tensor\_scatter\_nd\_add}. It is obvious that argument matching fails to synthesize the invocation code of \code{tf.tensor\_scatter\_nd\_add}.
In this case, a matching template can help \tech synthesize correct \invpat.
A \emph{matching template} is a code snippet elaborating a \mapipr. It presents an invocation of the \tgtapi, whose inputs are obtained from the arguments of the invocation of \srcapi.
Figure~\ref{fig:templatedoc} shows an example matching template (highlighted with underline) from the documentation of \code{tf.scatter\_nd}. It suggests that invoking \code{tf.tensor\_scatter\_nd\_add} is equivalent to invoking \code{tf.scatter\_nd} with proper argument mappings as shown in Figure~\ref{fig:matpat}.

To automatically detect such templates, \tech{} examines each code block in the documentation. Whenever a code snippet contains the invocation of another API, \tech{} extracts it as a potential candidate.
Note that not every \apipr has such templates. If \tech{} failed to extract any, the matching template will simply be \code{None}.
For \apiprs with a matching template, regardless of whether it is in top-$\TopK$ similar pairs, \tech synthesizes additional \invpat using the matching template as the target API invocation code.

\subsection{\verifier}
\label{subsec:verifier}

In this phase, \tech runs the invocation code synthesized for each matched API pair over a set of \emph{\verinp{}} to validate properties \propa{} and \propb{} (defined in Section~\ref{subsec:preliminary}). The \verinp{} are a collection of valid inputs for the \srcapi. These inputs can be collected from documentations, library tests, and existing DL models, so they are representative of the valid input space of the \srcapi. \revision{For example, in our implementation, we leverage state-of-the-art \freefuzz~\cite{freefuzz}, which can collect all such information fully automatically.
 }
 
\parabf{\propa}. \tech first checks whether the \invpat holds the \propa property given the \verinp. If so, \tech{} accepts this matching pattern and marks it as \propa{}. For example, the \invpat in Figure~\ref{fig:matpat} holds this property. 
For all \verinp{}, \code{tf.scatter\_nd} always have the same output with  \code{tf.tensor\_scatter\_nd\_add}. Hence, this API pair is labeled as \propa.

\parabf{\propb{}}. If the \invpat violates \propa{}, \tech checks whether the \srcapi and \tgtapi in the \invpat have the same status given \verinp. If so, \tech{} accepts this API pair and marks it as \propb{}. For instance, the \invpat shown in Figure~\ref{fig:statusmatch} does not hold the \propa{} property since the output of \code{AdaptiveAvgPool3d} is different from \code{AdaptiveMaxPool3d} over the input set. However, they have the same status over the \verinp{}, which allows them be labeled as \propb{}.

If the API pair is verified as \propa{} or \propb{}, the API pair that is accepted by the \verifier will be further tested in the next fuzzing phase; otherwise, \tech{} rejects this pair. Please note that it is crucial to have a set of representative \verinp. If the \verinp are not representative, the \verifier could mistakenly accept a wrong API relation when the \verinp{} do not cover certain important regions of the possible input space. We will study such false positives in detail in our experimental study (Section~\ref{sec:fpr}). On the other hand, the \verifier may also reject some true \mapiprs if the \verinp directly trigger real consistency bugs in this phase. Although it is hard to avoid such false negatives, our experimental results show that \tech detects \NumAllBugs bugs for popular DL libraries fully automatically, demonstrating the effectiveness of this design.

\subsection{\fuzzer}
\label{subsec:fuzzing}
In the last phase, \tech{} further leverages the verified API pairs to detect bugs for DL libraries. Specifically, \tech{} applies off-the-shelf mutation-based fuzzing techniques~\cite{freefuzz} to mutate the \srcapi inputs for generating diverse inputs for differential testing. 
For \propa{}, the source and target APIs are expected to have the same output. We can detect potential consistency bugs by comparing the detailed results of the source and target APIs.
For \propb{}, the source and target APIs are expected to have the same status. Therefore, we can detect bugs by comparing their statuses after execution.

\subsection{The Iterative Process}
\label{subsec:loop}
In order to cover more APIs by \mapipr{s}, \tech performs the above four phases iteratively until the fixed point or a given number of iterations $\IterI$ (by default $\IterI=10$ for this paper). In the \verifier phase, if the \tgtapi is not covered in the current iteration and its invocation generated by synthesizer has \code{Success} status, we will add this invocation into the database and label the \tgtapi as ``newly covered API''. After this iteration, if there is any newly covered API, \tech will re-run the framework \emph{with these newly covered APIs as the source APIs}; otherwise the fixed point has been reached, and \tech will terminate. \revision{It is worth mentioning that the entire iterative \tech approach is fully automated. For the newly covered APIs, the \verinp{} are also automatically borrowed from the source APIs' valid inputs.}

Figure ~\ref{fig:overview} presents one example for this iterative process. In the first iteration, the \verifier takes API pair \code{(torch.vsplit, torch.tensor\_split)} as input, and queries the invocation database \circled{1} to get a record for \code{torch.vsplit}. 
Invoking \code{torch.tensor\_split} \circled{2} results in \code{Success}. 
Assuming that \code{torch.tensor\_split} is not covered in the database at the beginning of the iteration, this successful invocation is then inserted into the database \circled{3}. 
In the next iteration, this invocation record will be retrieved \circled{1} to verify the matched API pairs with \code{torch.tensor\_split} as the \srcapi.

\section{Experimental Setup}

In the experiments, we address the following research questions:
\begin{itemize}
    \item RQ1: How effective is \tech in terms of API coverage?
    \item RQ2: What is the false positive rate of \tech?
    \item RQ3: How do different configurations affect \tech? 
    \item RQ4: Can \tech detect real-world bugs?
\end{itemize}

Our experiments are performed on \pt 1.10~\cite{paszke2019pytorch} and \tf 2.7~\cite{abadi2016tensorflow}, the latest stable release versions for the two most popular DL libraries, with 54K and 162K stars on GitHub. They have also been the most widely studied DL libraries in prior DL library testing work~\cite{cradle,lemon,freefuzz,guo2020audee}. The machine for running the experiments is equipped with 8-core 2.20GHz Intel Xeon CPU, 16GB RAM, Ubuntu 20.04, and Python 3.9.

\subsection{Implementation}

\parabf{\apirelfinder}. This phase provides candidates of \mapiprs for later verification and fuzzing. To find high-quality \mapiprs, we first use the bs4 Python package ~\cite{bs4_website} to parse the documentations of all \NumTotalAPI APIs from \tf and \pt. We collect both API signatures and descriptions from the documentation. To compute the TF-IDF embedding, we use the Snowball stemmer~\cite{porter2001snowball} to convert tokens into word stems. To compute the document embedding, we use the SentenceTransformer Python package~\cite{sbert_package_website} and use the pretrained model \code{all-MiniLM-L6-v2} as our $\sbert$. 

\parabf{\progsyn}. For argument matching, we use the munkres Python package~\cite{munkres_package_website} (which implements the Kuhn–Munkres algorithm) to solve the maximum weighted bipartite matching problem. For template matching, we automatically search and extract the code snippets for matching template from the documentation.

\parabf{\verifier}. We verify each \invpat with the help of \verinp. We obtain the valid inputs traced from various input sources used by state-of-the-art \freefuzz~\cite{freefuzz}, which include documentations, developer tests, and 202 DL models in the wild, and are representative to verify the function of APIs. We feed the first $100$ valid invocations of the \srcapi from the \freefuzz database into both the \srcapi and \tgtapi as the verifying inputs, and check if they have consistent behaviors. If there are fewer than $100$ records for the \srcapi, we use all of them.

\parabf{\fuzzer}. We leverage the fuzzing strategies of \freefuzz to mutate all the valid inputs traced for each \srcapi (within the \freefuzz database), and run all the generated inputs (1000 for each source API following the default setting of \freefuzz~\cite{freefuzz}) on both the \srcapi and \tgtapi for detecting consistency bugs.

\subsection{Metrics}

\parabf{\# of Covered API}. Following prior work in DL library testing ~\cite{freefuzz}, we report the number of covered APIs. 
An API is covered by \tech{} if it is successfully invoked by \verifier{} either as a \srcapi or a \tgtapi (i.e., invocations with the \success status). Since DL libraries contain a large number of APIs, API coverage is an important metric of test adequacy.

\parabf{False Positive Rate}. 
If an API pair satisfies that 1) at least one of its \invpats is accepted by the \verifier, and 2) at least one of its accepted \invpats is against its labeled property during the fuzzing phase, it is named an inconsistent API pair. 
False positive rate for inconsistent API pairs is the proportion of detected inconsistent API pairs which are false alarms. It is commonly used in prior work on fuzzing or testing~\cite{su2021fully, you2019profuzzer, donaldson2017automated}.

\parabf{\# of Detected Bugs. } Bug finding is the ultimate goal for fuzzing, and thus we also report the number of distinct bugs \tech{} finds.

\section{Result Analysis}
\label{sec:resultanalysis}

\subsection{RQ1: Effectiveness in API Coverage}

In this RQ, we aim to study the effectiveness of \tech{} in terms of covering more APIs with API relations. Table ~\ref{tab:apicov} shows the number of DL library APIs covered by  \tech{} and state-of-the-art \freefuzz{}. Column ``\#Total'' presents the total number of APIs in DL libraries, while Column ``Improvement'' presents the improvement of \tech over \freefuzz{}.
In the fuzzing stage, \tech{} covers \NumStudyAPI APIs, which is a huge improvement (\NumPercentStudyAPImore) over \freefuzz{} that covers only \NumFreeFuzzAPI APIs. For example, there are totally \NumTotalAPIPt \pt APIs, and \tech{} can cover \NumStudyAPIPt APIs, 128\% more than \freefuzz{}. A large number of APIs are not covered by \freefuzz{} because they are less frequently used and not covered by any of the three sources of \freefuzz{}. Leveraging \ars, \tech{} can successfully invoke these APIs with their relational APIs' inputs. The huge API coverage improvement demonstrates the potential of \tech{}.

\begin{table}[!t]\centering
\caption{Comparison with \freefuzz on API coverage }
\scalebox{0.85}{
\begin{tabular}{lrrrrr}\toprule
&\#Total & \#\freefuzz & \#\tech{} & Improvement(\%)  \\\midrule
\pt & \NumTotalAPIPt &\NumFreeFuzzAPIPt &\NumStudyAPIPt & 601 (128\%) \\ 
\tf & \NumTotalAPITf &\NumFreeFuzzAPITf &\NumStudyAPITf & 1214 (176\%) \\\hline
Total & \NumTotalAPI &\NumFreeFuzzAPI &\NumStudyAPI & \NumStudyAPImore (157\%) \\
\bottomrule
\end{tabular}}
\label{tab:apicov}
\end{table}

\begin{table}[!t]\centering
\caption{Verified API pairs }
\scalebox{0.9}{
\begin{tabular}{lrrrr}\toprule
& \propa & \propb & Total\\\midrule
\pt & \NumVerifyEqPairPt &\NumVerifySimPairPt & \NumVerifyPairPt \\ 
\tf & \NumVerifyEqPairTf &\NumVerifySimPairTf & \NumVerifyPairTf \\\hline
Total & \NumVerifyEqPair &\NumVerifySimPair & \NumVerifyPair \\
\bottomrule
\end{tabular}}
\label{tab:apirelations}
\end{table}

 \begin{table}[htp!]\centering
\captionsetup{labelfont={bf,color=black}, font={color=black}}
\caption{\revision{Source distribution of inferred API pairs }}\label{tab:pairseednew}
\scalebox{0.8}
{
\arrayrulecolor{black}
\begin{tabular}{>{\color{black}}l|>{\color{black}}r|>{\color{black}}r>{\color{black}}r>{\color{black}}r}\toprule
\textbf{} & \diagbox{\textbf{Src}}{\textbf{Tgt}} &\textbf{Seed} &\textbf{New} \\\midrule
\multirow{2}{*}{\textbf{\pt}}  &\textbf{Seed} &\numSeedSeedPairPt &544 \\
&\textbf{New} &598 &2246 \\ \midrule
\multirow{2}{*}{\textbf{\tf}}  &\textbf{Seed} &\numSeedSeedPairTf &1633 \\
&\textbf{New} &830 &5568 \\
\bottomrule
\end{tabular}
}
\arrayrulecolor{black}
 \end{table}

Table~\ref{tab:apirelations} further shows the number of verified API pairs detected by \tech{}. Columns ``\propa'' and ``\propb'' present the number of value-equivalent and status-equivalent API pairs accepted by the \verifier respectively.
\verifier accepts \NumVerifyPair API pairs in total, showing that such API relations are common in DL libraries. On \pt, \tech{} accepts more status-equivalent API pairs (\NumVerifySimPairPt) than value-equivalent (\NumVerifyEqPairPt). The reason is that the latter relation is stricter than the former, and status-equivalent API pairs are more common. For example, in term of splitting a tensor, \pt provides a set of APIs: \code{torch.split}, \code{torch.tensor\_split}, \code{torch.vsplit} (splits the tensor vertically), and \code{torch.dsplit} (splits the tensor depthwise).
It is worth noting that \tf has much more APIs than \pt, and \tech{} detects more value-equivalent API pairs than status-equivalent ones on \tf. This is because \tf contains lots of APIs for compatibility and low level access operations and thus has higher functional overlap: (1) The \code{tf.compat} module~\cite{tfcompat} contains 2579 redundant APIs to support forwards and backwards compatibility across \tf versions (e.g., v1 and v2). For example, \code{tf.compat.v1.layers.conv2d} is an alias for \code{tf.layers.conv2d}, and it allows user to use the \code{conv2d} layer with \tf v1 behaviour in \tf v2; (2) The \code{tf.raw\_ops} module~\cite{tfrawops} contains 1339 low level APIs to provide direct access to all \tf ops. For example, \code{tf.raw\_ops.Pad} adds padding to tensors and is a low level API compared to the high-level API \code{tf.pad} with the same functionality.

\revision{Table~\ref{tab:pairseednew} further presents a detailed distribution of the API pairs inferred by \tech based on whether the source/target APIs involve newly covered APIs. Column ``Src'' and Row ``Tgt'' present the categorization of the source and target APIs, respectively; Columns/Rows ``Seed'' and ``New'' indicate whether an API is from ``seed APIs'' covered by \freefuzz or is newly covered by \tech. Out of all the \NumVerifyPair API pairs verified by \tech{}, \NumVerifyPairSeedSeed (\numSeedSeedPairPt + \numSeedSeedPairTf) only involve APIs covered by the original \freefuzz, and all the remaining \NumVerifyPairNewNeeded pairs involve newly covered APIs, which shows the importance of leveraging API relations to cover more APIs. }

We also conduct a manual study to investigate why there are so many value-equivalent API pairs. Note that we do not look into status-equivalent API pairs because they are more intuitive (e.g., many APIs may share similar input parameter types and/or output behaviors). Since the number of verified value-equivalent API Pairs is huge, we select the set of equivalent API pairs explicitly specified in the documentations for our study.  We mine the documentations for all \NumTotalAPI \pt and \tf APIs to extract API pairs when one API explicitly reference another API in the API documentation. In this way, we automatically extract \NumDocAR{} API pairs and we manually categorize them in terms of why such relational APIs exist. 
\NumDocEqAR{} out of \NumDocAR{} API pairs are value-equivalent pairs. Note 1828 of them are backward compatibility pairs, and are unique to \tf. Thus, we group the remaining pairs into 5 main reason categories as shown in Table ~\ref{tab:equivreason}. Ease of programming is the main reason for \propa API pairs in both DL libraries. For example, \code{torch.det} is an alias for \code{torch.linalg.det}, and users can use the two symbols interchangeably. 
\revision{
Meanwhile, Deprecation is one of the minor reasons. It is worth noting that we include the deprecated APIs in our study since they are still in the code base and can help cover more new APIs as well as find potential consistency bugs.
}
Table~\ref{tab:equivreason} also provides examples to demonstrate each reason, where Column ``Example ($\apione$, $\apitwo$)'' refers to the (\srcapi, \tgtapi) pair. The last column explains the difference between the relational APIs.  For the Performance example, \code{tf.stack} and \code{tf.parallel\_stack} are equivalent APIs which pile a list of tensor up. \code{parallel\_stack} is more efficient than \code{stack} as the documentation of \code{tf.parallel\_stack} says ``\code{parallel\_stack} \emph{will copy pieces of the input into the output as they become available, in some situations this can provide a performance benefit.}''~\cite{parallelstack_website_tf}.   

\definecolor{Gray}{gray}{0.9}
\newcommand{\SrcAPITable}{$\apione$\xspace}
\newcommand{\TgtAPITable}{$\apitwo$\xspace}

\begin{table*}\centering
\caption{Classification of reasons for \propa API pairs}\label{tab:equivreason}
\scalebox{0.9}{
\begin{tabular}{l|rr|rr|lll}\toprule
\textbf{Reason} &\textbf{\# TF} &\textbf{\% TF} &\textbf{\# PT} &\textbf{\% PT} &\textbf{Example (\SrcAPITable, \TgtAPITable)} &\textbf{Diff. between \SrcAPITable and \TgtAPITable} \\\midrule
\textbf{Ease of programming} &460 &95.44\% &349 &91.36\% &\code{(torch.det, torch.linalg.det)} &\SrcAPITable is an alias for \TgtAPITable \\
\textbf{Performance} &10 &2.07\% &9 &2.36\% &\code{(tf.parallel\_stack, tf.stack)} &\SrcAPITable uses parallelism for efficiency. \\
\textbf{Special cases} &10 &2.07\% &10 &2.62\% &\code{(tf.boolean\_mask, tf.ragged.boolean\_mask)} &\TgtAPITable extends \SrcAPITable to ragged tensors. \\
\textbf{Numerical stability} &1 &0.21\% &4 &1.05\% &\code{(torch.linalg.inv, torch.linalg.solve)} &\TgtAPITable is faster and more numerically stable. \\
\textbf{Deprecation} &1 &0.21\% &10 &2.62\% &\code{(torch.qr, torch.linalg.qr)} &\SrcAPITable is deprecated. \\ 
\bottomrule
\end{tabular}}
\end{table*}

\subsection{RQ2: False Positives }
\label{sec:fpr}

False Positive Rate (FPR) is a common metric to evaluate the effectiveness of fuzz testing. Table~\ref{tab:bugfpr} shows the FPR of \tech on the DL libraries. We analyze all inconsistencies reported by the \fuzzer. Column ``All'' presents the total number of inconsistencies detected by \fuzzer{}. 
``TP'' (True Positive) means the number of true inconsistencies, and ``FP'' (False Positive) means the number of false alarms  (e.g., inconsistencies due to incorrectly inferred matched API pairs). We separately report the statistics of the \propa{} and \propb{} oracles, while the ``Overall'' statistics report the merged results.

The FPR of \tech{} is only \fpRate, which implies that our \ar detection and verification techniques are effective. 
The false positives mainly originate from incorrect \ar verification. The \verifier leverages valid inputs of the \srcapi to decide whether an \ar candidate is correct or not. However, the valid inputs come from the \freefuzz database and are not complete. As a result, the \verifier can mistakenly accept a wrong \ar when the set of inputs does not cover certain part of the possible input space and thus is insufficient to distinguish the different behaviors of the \ars. We can also observe that the FPR of \propa is lower than \propb. This is because that \propa is stricter than \propb, which makes it easier for the \verifier to reject wrong API relations over \verinp.

\begin{table}\centering
\caption{False positive rate of \tech}
\label{tab:bugfpr}
\scalebox{0.85}{
\begin{tabular}{llrrrrr}\toprule
\textbf{} & \textbf{Oracle} & \textbf{\# All} & \textbf{TP} &\textbf{FP} &\textbf{FPR} \\\midrule
 & \textbf{\propa{}} & 412 & 364 & 48 & 11.65\% \\
\textbf{\pt} & \textbf{\propb{}} & 853 & 554 & 299 & 35.05\% \\
 & \textbf{Overall} & 1265 & 918 & 347 & \fpRatePt \\ \hline
 & \textbf{\propa{}} & 97 & 68 & 29 & 29.90\% \\
\textbf{\tf} & \textbf{\propb{}} & 363 & 209 & 154 & 42.42\% \\
 & \textbf{Overall} & 460 & 277 & 183 & \fpRateTf \\ \hline
 & \textbf{\propa{}} & 509 & 432 & 77 & 15.13\% \\
\textbf{Total}& \textbf{\propb{}} & 1216 & 763 & 453 & 37.25\% \\
 & \textbf{Overall} & 1725 & 1195 & 530 & \fpRate \\
\bottomrule
\end{tabular}}
\end{table}

\subsection{RQ3: Impacts of Configuration}
\label{sec:rq3}

\begin{figure*}
    \centering
    \captionsetup{justification=centering}
    \begin{subfigure}{0.30\textwidth}
        \centering
        \includegraphics[width=\textwidth]{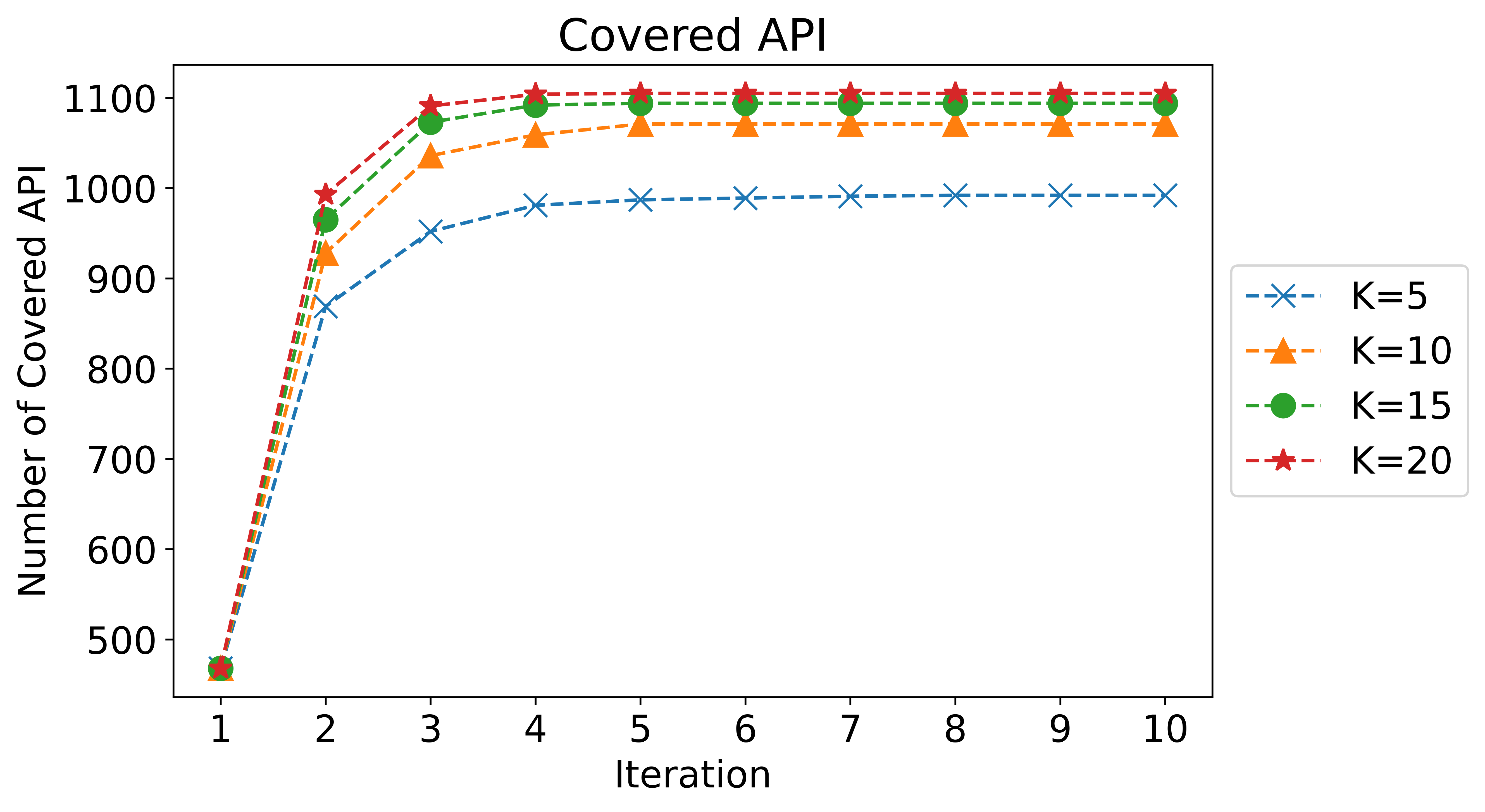}
        \caption{Number of covered APIs}
         \label{fig:trendapi}
    \end{subfigure}\hfill
    \begin{subfigure}{0.30\textwidth}
        \centering
        \includegraphics[width=\textwidth]{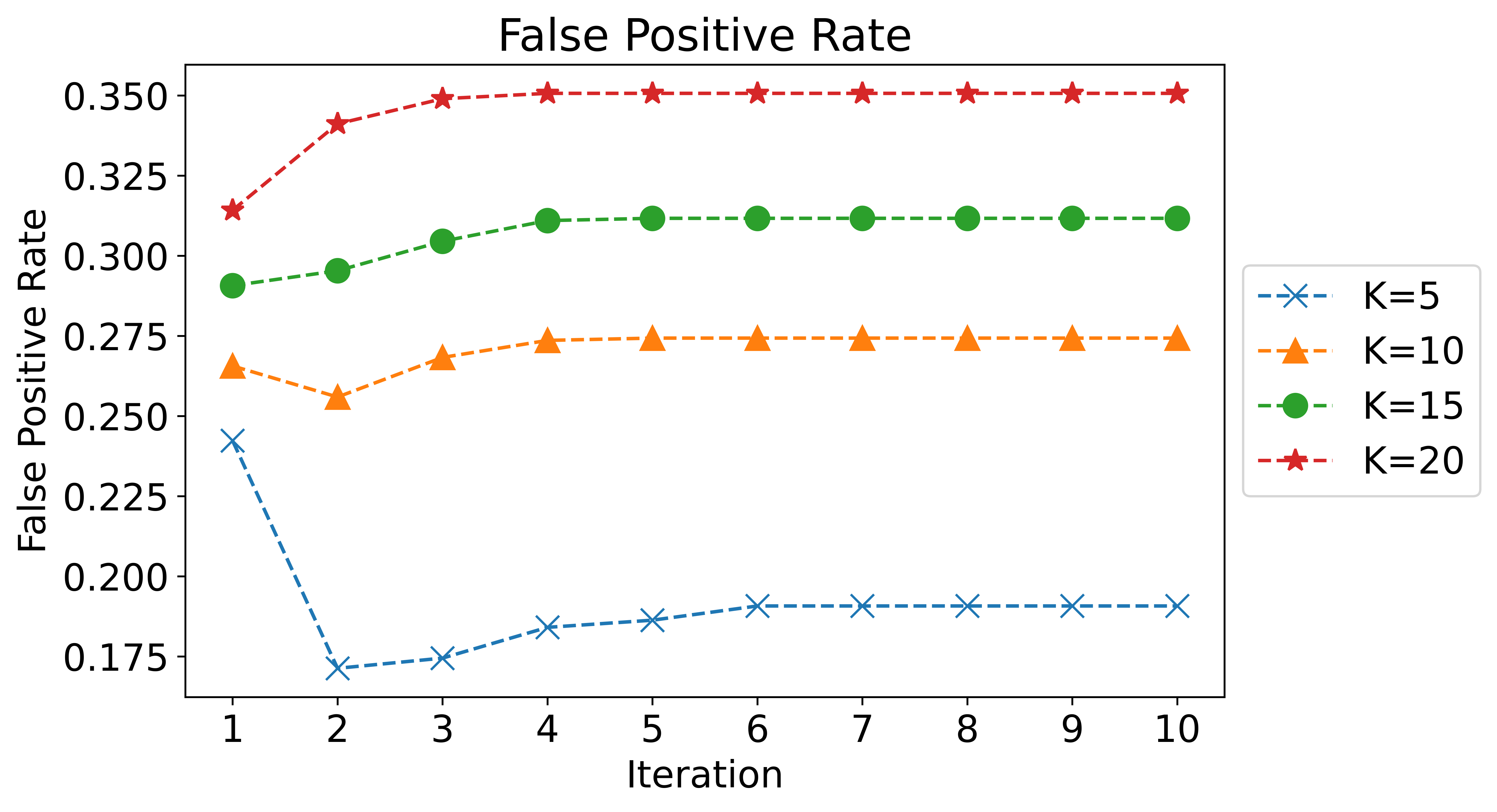}
        \caption{False positive rate}
        \label{fig:trendstatus}
    \end{subfigure}\hfill
    \begin{subfigure}{0.30\textwidth}
        \centering
        \includegraphics[width=\textwidth]{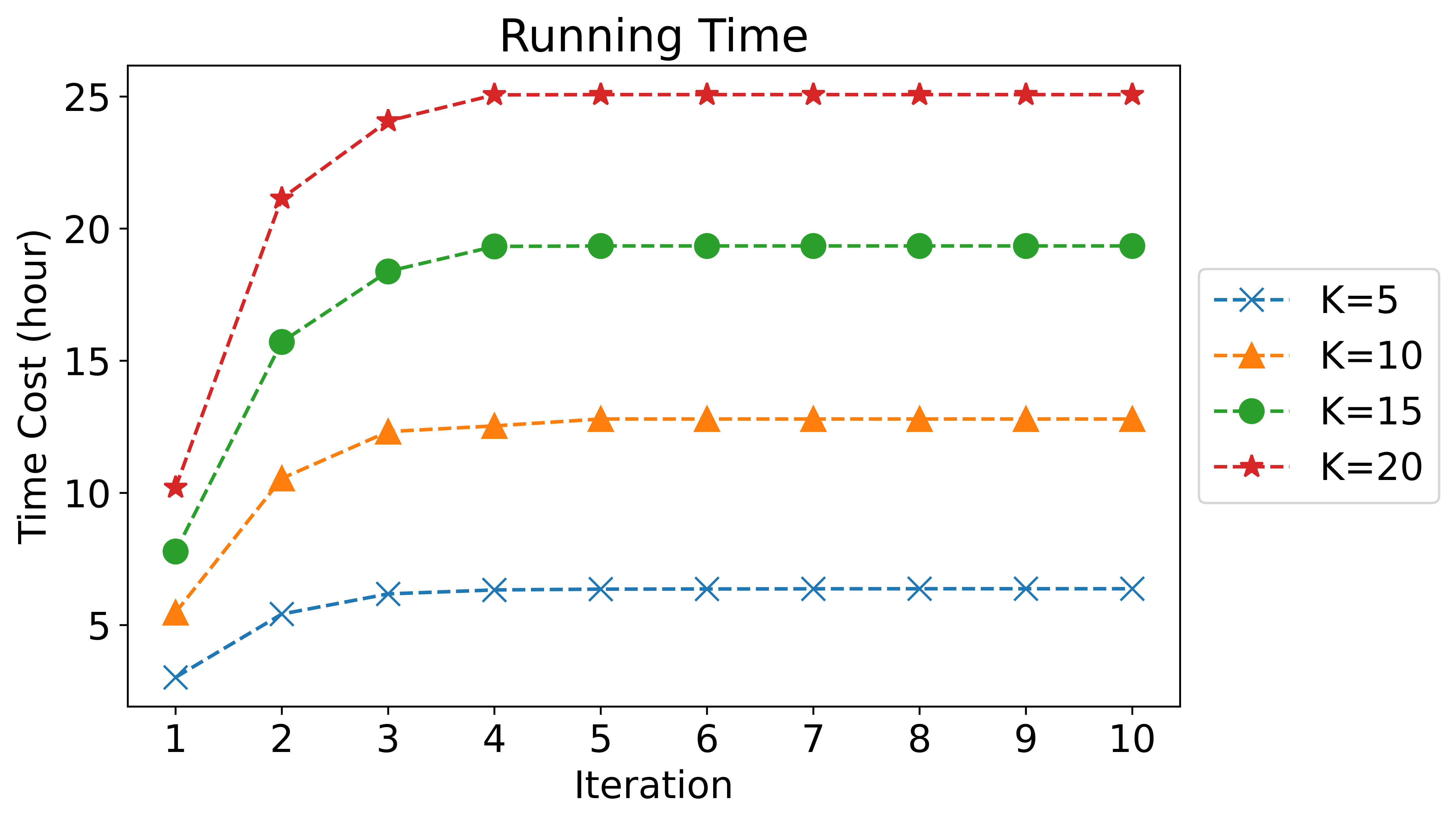}
        \caption{Running time}
        \label{fig:trendfpr}
    \end{subfigure}
        \caption{Analysis of hyper-parameter top-$K$ and iteration $\IterI$ on \pt }
        \label{fig:ktrendpt}
\end{figure*}

\begin{figure*}
    \centering
    \captionsetup{justification=centering}
    \begin{subfigure}{0.30\textwidth}
        \centering
        \includegraphics[width=\textwidth]{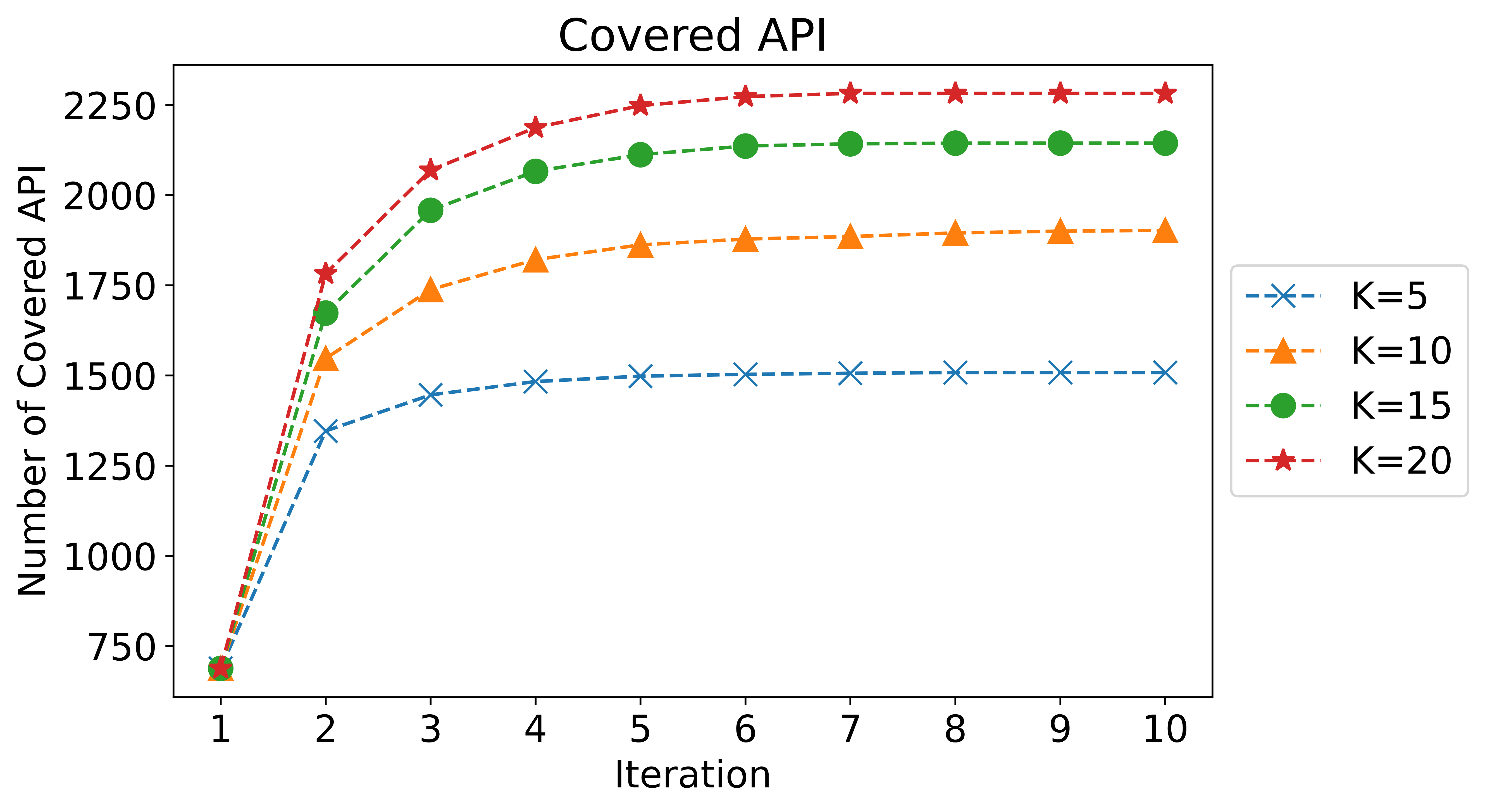}
        \caption{Number of covered APIs}
        \label{fig:tftrendapi}
    \end{subfigure}\hfill
    \begin{subfigure}{0.30\textwidth}
        \centering
        \includegraphics[width=\textwidth]{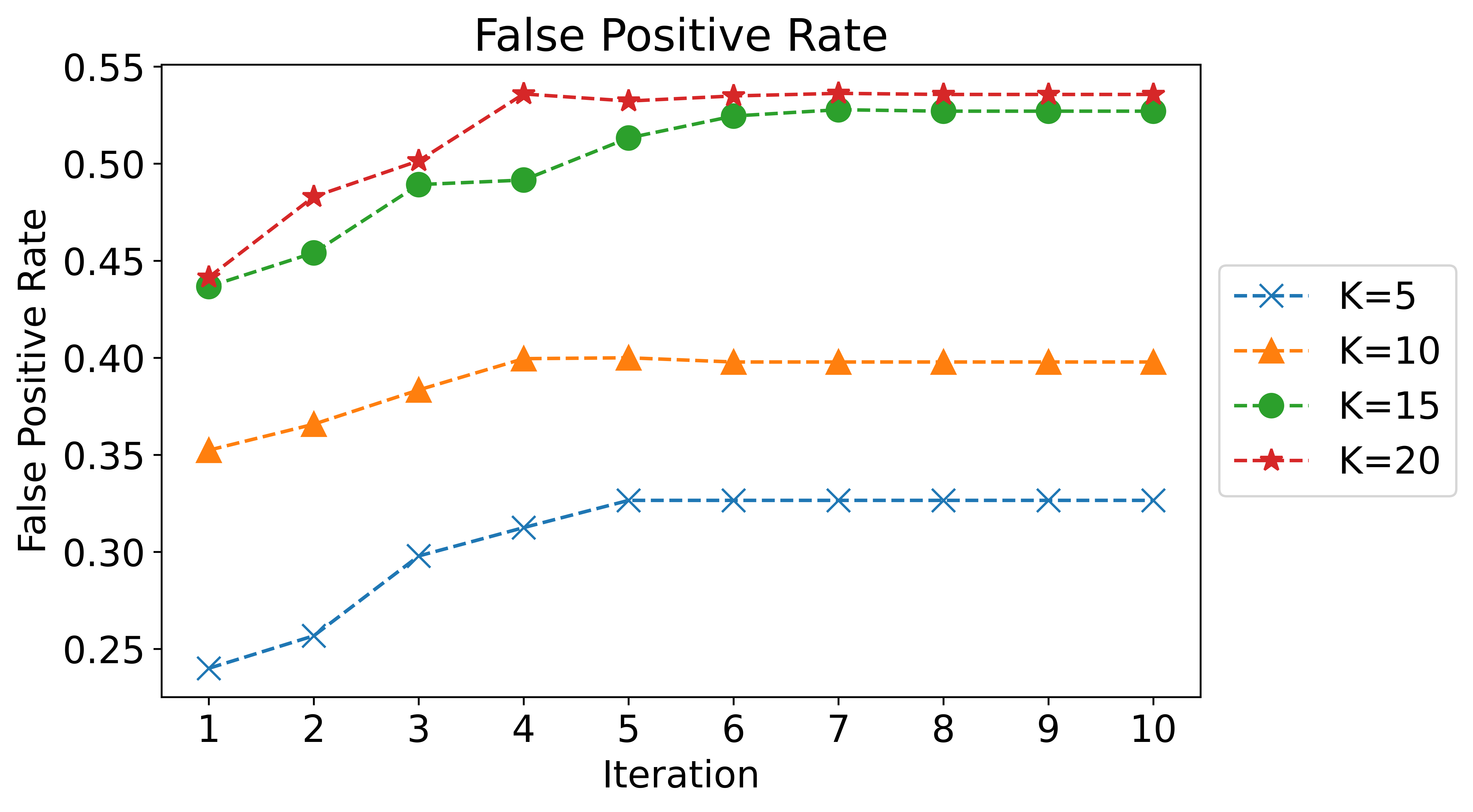}
        \caption{False positive rate}
        \label{fig:tftrendstatus}
     \end{subfigure}\hfill
     \begin{subfigure}{0.30\textwidth}
        \centering
        \includegraphics[width=\textwidth]{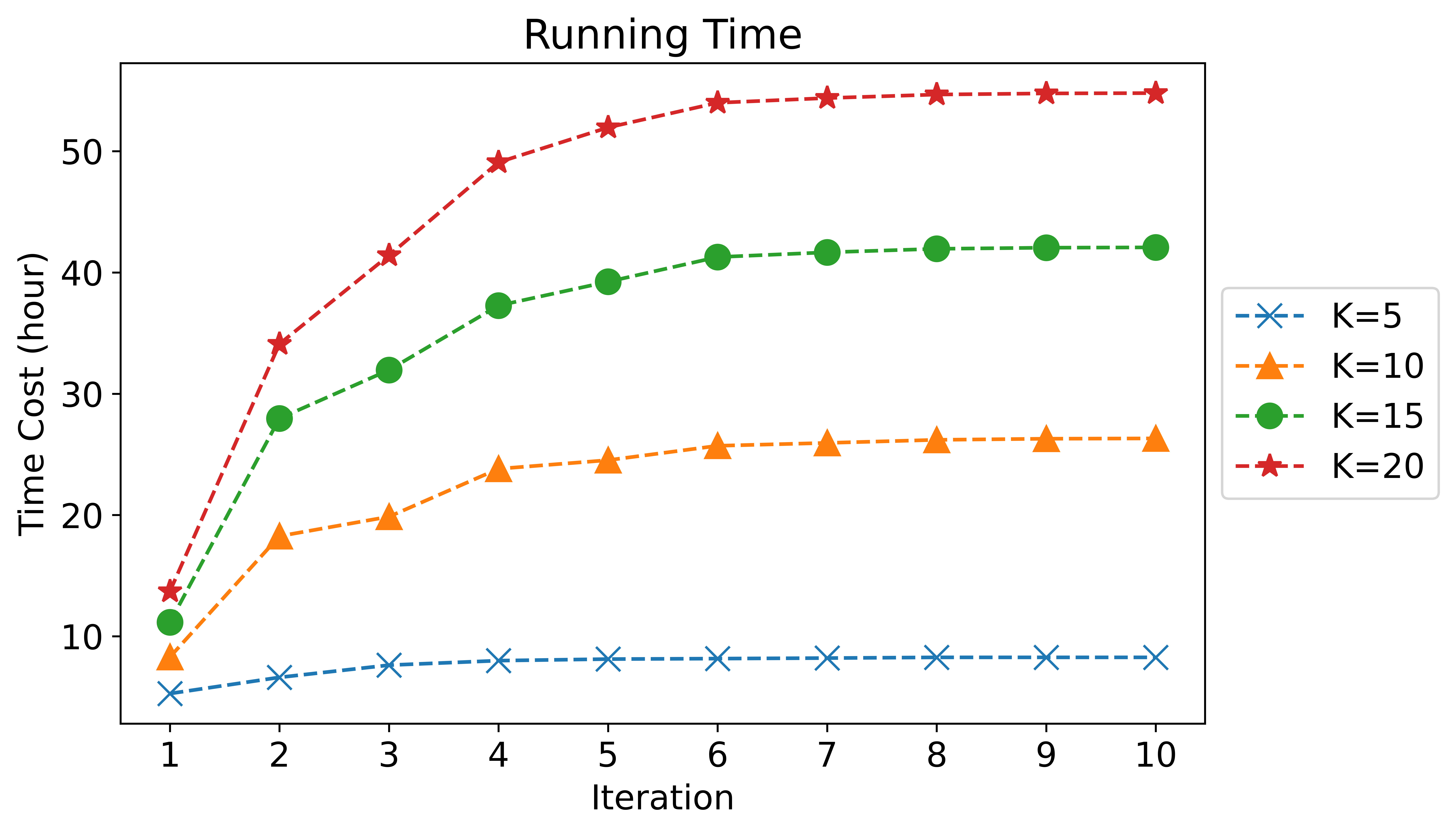}
        \caption{Running time}
        \label{fig:tftime}
    \end{subfigure}
        \caption{Analysis of hyper-parameter top-$K$ and iteration $\IterI$ on \tf}
        \label{fig:ktrendtf}
\end{figure*}

In this RQ we analyze how different configurations affect the performance of \tech{}, including API coverage, False Positive Rate (FPR), and the running time. We focus on two hyper-parameters, $\TopK$ (the number of matched pairs for each source API, discussed in Section~\ref{subsec:apirelafinder}) and $\IterI$ (the number of iterations, discussed in Section~\ref{subsec:loop}). The default values for $\TopK$ and $\IterI$ are both $10$ in \tech. To figure out the impact of $\TopK$, we run our experiments with different $\TopK$ values of $5,10,15,20$. We also run \tech{} for up to $10$ iterations and show the impact of different $\IterI$ values from 1 to 10.
Figures~\ref{fig:ktrendpt} and ~\ref{fig:ktrendtf} show the results under different configurations. The \emph{x} axis shows the iterations, and the \emph{y} axis presents the number of covered APIs, FPR, and the running time. The results for different $\TopK$ values are shown in different lines.  We can observe that \tech{} will terminate (i.e., reaching fixed points) in at most \NumIterTerminatePt iterations on \pt, and on \tf it either terminates in at most \NumIterTerminateTf iterations, or only covers 2 new APIs in the last (i.e. 10th) iteration. 

The impact of $\TopK$ is similar for \pt and \tf. First, the number of covered APIs of $\TopK=10,15,20$ are close, all significantly higher than $\TopK=5$. Second, the FPR increases as $\TopK$ increases. The reason is that less similar APIs can incur more false positives, indicating the effectiveness of our API similarity computation. Third, the running time is increasing approximately linearly with $\TopK$. Therefore, it is reasonable to set $\TopK$ as 10 in our default setting, as setting $\TopK$ to be higher than $10$ would bring marginal benefit in terms of API coverage, but degrade FPR and time cost. 

For the impact of $\IterI$, we have the following observations. First, as expected, the number of covered APIs increases at a lower and lower pace with the increase of $\IterI$, and converges within $10$ iterations. Second, the false positive rate generally increases with the iteration. The reason is that source APIs for later iterations are typically target APIs from earlier iterations, and may have less and less valid inputs (since they may fail on some inputs from the original source APIs) for verifying inferred API pairs. The only exception is that FPR drops when $\IterI$ increases to $2$ with $\TopK$ = $5$ or $10$ on \pt. We look into the data and observe that \tech{} is able to accidentally detect a large number of true positives in the $2^{nd}$ iteration (e.g., \code{torch.Tensor.*} APIs and their value-equivalent APIs \code{torch.*}).  Third, the running time increases more dramatically at early iterations and grows slowly in later iterations, which is consistent with the growth of the number of covered APIs. 

We can also observe that the total running time of the default \tech{} is 12.8h for \pt and 26.3h for \tf. Such cost is actually quite common for fuzzing techniques, e.g., various fuzzing techniques have been applied for 24h or even more~\cite{bohme2017coverage, you2019profuzzer,wen2020memlock,she2020mtfuzz,liu2022coverageguided} (including the recent \lemon work~\cite{lemon}). 

\begin{table*}[!htp]\centering
\caption{Summary of detected bugs\label{tab:allbugs}
}

\scalebox{0.9}{
\begin{tabular}{lrrr|>{\color{black}}r>{\color{black}}r>{\color{black}}r|>{\color{black}}r>{\color{black}}r>{\color{black}}r|>{\color{black}}r>{\color{black}}r>{\color{black}}r>{\color{black}}r}\toprule
\textbf{} &\multicolumn{3}{c}{\textbf{Total}} &\multicolumn{3}{>{\color{black}}c}{\textbf{Rejected}} &\multicolumn{3}{>{\color{black}}c}{\textbf{Confirmed}} &\multicolumn{3}{>{\color{black}}c}{\textbf{Fixed}} \\\cmidrule{2-13}
&\textbf{Total} &Value &Status &\textbf{Total} &Value &Status &\textbf{Total} &Value &Status &\textbf{Total} &Value &Status \\\midrule
\pt &\textbf{\NumAllBugsPt} &76 &45 &\textbf{\NumRejectBugPt} &3 &3 &\textbf{\NumPreviouslyUnknownConfirmPt} &33 &38 &\textbf{\numFixedPt} &9 &8 \\
\tf &\textbf{\NumAllBugsTf} &22 &19 &\textbf{\NumRejectBugTf} &0 &1 &\textbf{\NumPreviouslyUnknownConfirmTf} &22 &13 &\textbf{\numFixedTf} &3 &9 \\
Total &\textbf{\NumAllBugs} &98 &64 &\textbf{\NumRejectBug} &3 &4 &\textbf{\NumPreviouslyUnknownConfirm} &55 &51 &\textbf{\numFixedBugs} &12 &17 \\
\bottomrule
\end{tabular}
}
\end{table*}

\begin{table}[!htp]\centering
\captionsetup{labelfont={bf,color=black}, font={color=black}}
\caption{\revision{Source distribution of confirmed bugs}\label{tab: }}

\scalebox{0.85}{
\arrayrulecolor{black}
\begin{tabular}{>{\color{black}}l|>{\color{black}}r|>{\color{black}}r>{\color{black}}r>{\color{black}}r|>{\color{black}}r>{\color{black}}r>{\color{black}}r>{\color{black}}r}\toprule
& &\multicolumn{3}{>{\color{black}}c}{\textbf{Single API bugs}} &\multicolumn{3}{|>{\color{black}}c}{\textbf{Relational API bugs}} \\\cmidrule{3-8}
\textbf{} &\textbf{Total} &\textbf{Total} &\textbf{Seed} &\textbf{New} &\textbf{Total} &\textbf{Seed-only} &\textbf{Others} \\\midrule
\textbf{PT} &71 &18 &9 &9 &53 &26 &27 \\
\textbf{TF} &35 &1 &0 &1 &34 &9 &25 \\
\textbf{Total} &106 &19 &9 &10 &87 &35 &52 \\
\bottomrule
\end{tabular}
}
\arrayrulecolor{black}
\label{tab:bugsbreakdown}
\end{table}

\subsection{RQ4: Bugs Detected}
Table~\ref{tab:allbugs} presents the summary of real-world bugs detected by \tech{} for the two studied DL libraries. Column ``Total'' shows the total number of bugs detected by \tech{}, and Columns ``Value'' and ``Status'' show the number of bugs detected with the \propa and \propb oracle respectively. 
We also present the number of bugs rejected by developers (Column ``Rejected''), confirmed as previously unknown (Column ``Confirmed''), and the number of previously unknown bugs that have already been fixed (Column ``Fixed''). We can observe that \tech{} is able to detect \NumAllBugs bugs in total, with only 7 rejected by developers, \NumPreviouslyUnknownConfirm confirmed by developers as previously unknown bugs  (\numFixedBugs already fixed), and all others pending. 
Among those \NumPreviouslyUnknownConfirm confirmed bugs, only \NumFreeFuzzCanFindBugs can be found by \freefuzz, and none of them can be detected by \cradle, \audee, or \lemon.  
Furthermore, we have also found \NumDocBugPt documentation bugs for \pt and \NumDocBugTf for \tf during the experiment (these document bugs are not included in Table~\ref{tab:allbugs}).

\revision{The bugs detected by \tech{} can also be categorized into single API bugs\footnote{\revision{The single API bugs are unexpected crashes caused by single buggy APIs. In Table~\ref{tab:allbugs}, we categorize such bugs as ``Value''/``Status'' if they were detected when testing API pairs with the \propa/\propb oracle (although they can be detected without such relational API oracles).}} and consistency bugs between relational API pairs.   Table~\ref{tab:bugsbreakdown} shows the breakdown of the \NumPreviouslyUnknownConfirm confirmed bugs based on the two cases.}
\revision{ For the \numSingleAPI single API bugs detected by \tech{}, \numSeedAPI are in the "seed APIs" (Column ``Seed''), while \numNewAPI are bugs in the newly covered APIs (Column ``New''), emphasizing the importance of leveraging API relations to cover more APIs. Meanwhile, all the remaining \numConsistent ones are consistency bugs, demonstrating the effectiveness of using API relations as the oracle for DL library testing.
More specifically, out of the \numConsistent consistency bugs, \numConsistentSeedOnly are inconsistencies between “seed APIs” (Column ``Seed-only'') while \numConsistentNew involve newly covered APIs (Column ``Others''), further indicating that covering new APIs contributes to consistency bug detection.} 

Notably, \tech{} is able to detect \NumHighPrio high-priority bugs for \pt (note that \tf is not discussed here as it does not have such labels). These bugs are marked as ``high priority'' because they are critical and require urgent resolution. The reported ``high priority'' bugs are highly valued by \pt developers and have raised heated discussions.  During the three months from our first bug issue (November 23, 2021) to March 1, 2022, there have been a total of 170 high-priority issues in the entire \pt issue-tracking system~\cite{pytorchrepo}, of which we contributed 23 (\PortionHighPrio), emphasizing the effectiveness of \tech.

We next present some example bugs detected by \tech:

\parabf{Out-of-Bounds Read (\propa, \merged)} 
The bug shown in Figure~\ref{fig:torchkthvalue} is detected for API pair \CodeIn{torch.kthvalue(tensor, k)} and \CodeIn{tensor.kthvalue(k)} via \propa. The latter API is one of the methods from \CodeIn{torch.Tensor}~\cite{pttensor}, which is the fundamental class of \pt. Given the same tensor (\CodeIn{input}) and argument (\CodeIn{k}), the returned results (\CodeIn{result1} and \CodeIn{result2}) are not equal (i.e., assertion fails on Line 5).
After debugging, we find that the returned results can actually be different across runs, indicating that it reads values from memory locations outside of user-controlled data! This bug is a silent error, and has a severe security implication: without proper range checking of \CodeIn{k}, users may be able to read data outside of the allocated memory bounds (i.e., out-of-bound read). This bug is labeled as ``high priority'', and developers have fixed it immediately.

\begin{figure}
    \begin{minted}[mathescape, linenos, numbersep=5pt, gobble=0, fontsize=\footnotesize, frame=lines, framesep=2mm]{python}
input = torch.tensor([0,1,2,3,4])
k = 6
result1 = torch.kthvalue(input,k)
result2 = input.kthvalue(k)
torch.testing.assert_close(result1, result2) # fail
    \end{minted}
	\caption{Out-of-Bound Read in \CodeIn{torch.kthvalue}}
	\label{fig:torchkthvalue}
\end{figure}

\parabf{Inconsistent Check (\propb, \merged)}
Figure~\ref{fig:torchadaptive} shows a bug for \CodeIn{torch.nn.AdaptiveAvgPool3d} and \CodeIn{torch.nn.AdaptiveMaxPool3d} (found via \propb). With exactly the same input tensor (\CodeIn{tensor}) and argument (\CodeIn{output\_size}), \CodeIn{AdaptiveAvgPool3d} runs without exception (Line 4) while \CodeIn{AdaptiveMaxPool3d} throws \CodeIn{RuntimeError: Trying to create tensor with negative dimension} (Line 6). After inspection, we find that \CodeIn{AdaptiveAvgPool3d} lacks checking for dimensions with negative integers. This can be disastrous since users may inadvertently introduce a bug into their model, but no warning/exception is raised. This bug has also been fixed.

\begin{figure}
    \begin{minted}[mathescape, linenos, numbersep=5pt, gobble=0, fontsize=\footnotesize, frame=lines, framesep=2mm]{python}
output_size = [-36, 0, 0]
tensor = torch.rand([4, 4, 128, 2048, 4])
layer1 = torch.nn.AdaptiveAvgPool3d(output_size)
result1 = layer1(tensor) # no exception
layer2 = torch.nn.AdaptiveMaxPool3d(output_size)
result2 = layer2(tensor) # RuntimeError
    \end{minted}
	\caption{Inconsistent check for \CodeIn{torch.nn.AdaptiveAvgPool3d}}
	\label{fig:torchadaptive}
\end{figure}

\parabf{Wrong Computation (\propa, \rejected)}
The bug shown in Figure~\ref{fig:ptwrong} is detected by \propa oracle for API pair \CodeIn{torch.std\_mean} and \CodeIn{torch.mean}. Both of them could compute the mean of all elements in the \CodeIn{input} tensor. 
However, \tech has found that for certain input tensors, \CodeIn{torch.std\_mean} has different returned value with \CodeIn{torch.mean} (i.e., assertion fails on Line 5). Although we believe it is a bug, the developers said these two APIs are not expected to output the same mean and rejected this bug report. Note that the other \NumOtherRejectBug rejected bugs are similar (there are indeed inconsistencies, but developers feel unnecessary to fix).

\begin{figure}
    \begin{minted}[mathescape, linenos, numbersep=5pt, gobble=0, fontsize=\footnotesize, frame=lines, framesep=2mm]{python}
input = torch.tensor([[0.5786, 0.1719, 0.3760, 0.2939, 0.3984],
        [0.5361, 0.7104, 0.8765, 0.0903, 0.0483]], dtype=torch.float16)
result1 = torch.std_mean(input) # 0.4080
result2 = torch.mean(input) # 0.4082
torch.testing.assert_close(result1, result2) # fail
    \end{minted}
	\caption{Inconsistency for \CodeIn{torch.std\_mean} and \CodeIn{torch.mean}}
	\label{fig:ptwrong}
\end{figure}

\subsection{Threats to validity}

The threats to internal validity mainly lie in the correctness of the \tech implementation. To reduce such threats, the first three authors have performed extensive testing and code review of \tech. Moreover, we have released our code/scripts in our project website for public review~\cite{deeprelrepo}. The threats to external validity mainly lie in the evaluation benchmarks used. To demonstrate the generalizability of \tech, we have evaluated \tech on two widely used DL libraries, \tf and \pt. Last but not least, the threats to construct validity mainly lie in the metrics used. To reduce such threats, we adopt the number of covered APIs and detected bugs, following prior work on DL library testing~\cite{cradle,lemon,freefuzz}. Moreover, we have also included false positive analysis.

\section{Related Work}

We have talked about related work on DL library testing in Section~\ref{section:background}. Therefore, in this section, we mainly focus on existing work targeting API relations for other software systems.
API mappings refer to the process of finding equivalent APIs between different libraries or programming languages. Prior work~\cite{nguyen2016mapping,nguyen2017exploring} proposed to train neural networks to mine API mappings by learning transformations between the vector spaces of APIs from different languages (e.g., Java and C\#). Another work~\cite{bui2019towards} leveraged unsupervised domain adaptation approach to automatically construct and align vector spaces for identifying API mappings with much less human efforts. Our work is different from all such work in that \tech focuses on fuzzing while aforementioned work targets code migration. 
Also, \tech utilizes not only equivalent APIs, but also other relational APIs for fuzzing. Moreover, \tech obtains relational APIs purely from documentation, and then verifies the API relations dynamically; thus, \tech does not need extensive API usages or training data of existing API pairs (which can be hard to obtain). 

Function synonyms~\cite{defreez2018path} are functions that play a similar (\emph{not} necessarily semantically equivalent) role in code. \ucklee~\cite{ramos2011practical} leverages symbolic execution to check two different implementations/versions of the same function in C open source libraries. Func2vec~\cite{defreez2018path} is a technique for finding function synonyms in Linux file systems and drivers via learning function embeddings. Func2vec trains a neural network on sentences generated using random walks of the interprocedural control-flow graph of the program, and it is applied to improve the quality of error handling specifications for Linux code. More recently,
\fpdiff~\cite{vanover2020discovering} aims to automatically identify synonymous functions across multiple numerical libraries, and performs differential testing to detect discrepancies of these function synonyms. While \fpdiff is closely related, it mainly considers synonymous APIs with same functionalities and argument lists across different libraries and is typically applied when there exist libraries with close design; in contrast, \tech is more general -- it considers arbitrary relational APIs with value/status equivalence within \emph{any given} library. Also, \fpdiff can only target numerical library APIs taking \texttt{double} or \texttt{int} variables as input, whereas the input of APIs in DL libraries can be much more complex (e.g., tensors of different dimensions/types, objects, strings, etc.). Moreover, \fpdiff is a purely dynamic technique that finds synonymous functions via running them on an \emph{elementary} set of \texttt{double/int} values, while it is impossible to find such an universal elementary input set for DL library APIs. Lastly, our study has found that DL library documentations contain various valuable information and shown for the first time that there can be plenty of equivalent/similar APIs within a given DL library, which can substantially help with fuzzing DL libraries (and beyond).

\revision{
Throughout this paper, we have viewed each API as a test object, and thus our technique of fuzzing relational API pairs falls into the differential testing category (following \fpdiff~\cite{vanover2020discovering}). 
Meanwhile, if we treat the entire systems under test (e.g., \tf or \pt) as test objects, this work can also be viewed as an instance/extension of the metamorphic oracle generation work, which aims to automatically infer metamorphic relations as test oracle~\cite{goffi2014search,mattavelli2015synthesis,memo,kanewala2014techniques,zhang2014search,troya2018automated,xiang2019genetic,zhang2019automatic}. For example, \sbes~\cite{goffi2014search,mattavelli2015synthesis} synthesizes sequences of method invocations that are equivalent to a target method based on the observed dynamic program behaviors. The recent state-of-the-art technique \memo~\cite{memo} mainly targets mature open-source Java projects and automatically derives metamorphic relations from natural language specifications.
 Meanwhile, \memo can only infer metamorphic relations that are well documented and lacks invocation synthesizer or API relation verifier; thus the effectiveness of \memo strictly depends on documentation completeness, correctness, and preciseness. In fact, very few metamorphic relations in DL libraries are explicitly described with informative sentences (with code templates) and can be directly translated into valid oracles. } \revision{We substituted the first three phases of \tech{} with \memo, and found that it can infer only 17/28 API relations for \tf/\pt (compared to 8808/4290 relations found by \tech), and can at most detect 8 (including 6 found by \memo and 2 found by \freefuzz on the APIs involved in the \memo inferred relations) of the 106 confirmed bugs detected by \tech.}

\section{Conclusion}

We have introduced \tech{}, the first fully automated end-to-end approach to \revision{fuzzing DL libraries via inferring relational APIs}. \tech{} can ``borrow'' test inputs from any API to test its relational APIs, and can leverage relational APIs as reference implementations for performing differential testing. The extensive study of \tech{} on \pt and \tf shows that \tech{} is able to detect \NumAllBugs{} bugs in total, with \NumPreviouslyUnknownConfirm already confirmed by the developers as previously unknown bugs. Notably, \tech has detected \PortionHighPrio of the high-priority bugs for the entire \pt issue-tracking system in a three-month period. Also, besides the \NumAllBugs{} code bugs, we were also able to detect \NumDocBug documentation bugs (all confirmed).

\bibliographystyle{abbrv}
\bibliography{main}

\end{document}